\documentclass[sn-mathphys,Numbered]{sn-jnl}%

\usepackage{graphicx}%
\usepackage{multirow}%
\usepackage{amsmath,amssymb,amsfonts}%
\usepackage{amsthm}%
\usepackage{mathrsfs}%
\usepackage[title]{appendix}%
\usepackage{xcolor}%
\usepackage{textcomp}%
\usepackage{manyfoot}%
\usepackage{booktabs}%
\usepackage{algorithm}%
\usepackage{algorithmicx}%
\usepackage{algpseudocode}%
\usepackage{listings}%
\hypersetup{hypertexnames=false}

\usepackage{siunitx}
\usepackage{gensymb}
\usepackage{mhchem}

\usepackage{cleveref}
\crefname{figure}{Fig.}{Figs.}
\crefname{equation}{Eq.}{Eqs.}
\creflabelformat{methods}{#2Methods#3}
\crefname{methods}{}{}
\crefname{supsection}{Supplementary}{Supplementary}
\crefname{supfigure}{Supplementary Fig.}{Supplementary Figs.}
\crefname{supvideo}{Supplementary Video}{Supplementary Videos}

\let\vec\mathbf

\begin{document}

\title[Clocked dynamics in artificial spin ice]{Clocked dynamics in artificial spin ice}

\author*[1]{\fnm{Johannes H.} \sur{Jensen}}
\email{johannes.jensen@ntnu.no}
\equalcont{These authors contributed equally to this work.}
 
\author*[2]{\fnm{Anders} \sur{Str\o mberg}}
\email{anders.stromberg@ntnu.no}
\equalcont{These authors contributed equally to this work.}

\author[2]{\fnm{Ida} \sur{Breivik}}

\author[1]{\fnm{Arthur} \sur{Penty}}
\author[3]{\fnm{Miguel Angel} \sur{Ni\~no}}
\author[3]{\fnm{Muhammad Waqas} \sur{Khaliq}}
\author[3]{\fnm{Michael} \sur{Foerster}}
\author[1]{\fnm{Gunnar} \sur{Tufte}}
\author[2]{\fnm{Erik} \sur{Folven}}

\affil*[1]{\orgdiv{Department of Computer Science}, \orgname{Norwegian University of Science and Technology}, Trondheim, Norway}

\affil*[2]{\orgdiv{Department of Electronic Systems}, \orgname{Norwegian University of Science and Technology}}

\affil[3]{\orgname{ALBA Synchrotron Light Facility}, \orgaddress{\street{Carrer de la Llum 2 -- 26}, \city{Cerdanyola del Vallés}, \postcode{08290}, \state{Barcelona}, \country{Spain}}}

\abstract{ %
Artificial spin ice (ASI) are nanomagnetic metamaterials exhibiting a wide range of emergent properties, which have recently shown promise for neuromorphic computing.
However, the lack of efficient protocols to control the state evolution of these metamaterials has been limiting progress.
To overcome this barrier, we introduce \emph{astroid clocking}, a global field protocol offering discrete, gradual evolution of spin states.
The method exploits the intrinsic switching astroids and dipolar interactions of the nanomagnets to selectively address ASI spins in sequence.
We demonstrate, experimentally and in simulations, how astroid clocking of pinwheel ASI allows ferromagnetic domains to be gradually grown or reversed at will.
More complex dynamics arise when the clock protocol allows both growth and reversal to occur simultaneously.
Astroid clocking offers unprecedented control and understanding of ASI dynamics in both time and space, extending what is possible in nanomagnetic metamaterials.

}


\maketitle

\section{Controlling artificial spin ice}
Artificial spin ice (ASI) are systems of coupled nanomagnets arranged on a two-dimensional lattice.
The nanomagnets are elongated, giving them two stable magnetization directions, thus behaving as artificial spins.
Dipolar interactions give rise to a rich variety of emergent behavior, as determined by the ASI geometry\citep{Ladak2010,Gliga2017,Sendetskyi2019}.
As this behavior can be probed directly, ASIs have attracted considerable interest as model systems for the study of fundamental physics\citep{Heyderman2013,Skjervo2020}.
More recently, ASIs have shown promise as substrates for computation\citep{Jensen2018,Arava2018,Gypens2018,Jensen2020,Hon2021,Kaffash2021,Gartside2022}.

External fields are the primary method used to perturb ASIs in a controlled manner.
Various global field protocols have been employed.
For example, a cycled in-plane field is often used to characterize magnetization reversal\citep{Ladak2010,Schumann2010,Mengotti2011,Morgan2011,Pollard2012,Leon2013,Gilbert2015,Li2018,Bingham2022,BegumPopy2022}.
Another approach is to use a rotating field with slowly decreasing amplitude to effectively anneal the ASI to a low energy state\citep{Wang2006,Wang2007,Nisoli2007,Ke2008,Qi2008,Nisoli2010,Phatak2011,Perrin2016}.
While there are variations of these simple field protocols, more complex protocols are largely unexplored.

These approaches use field strength to modulate ASI behavior, which will typically result in uncontrolled avalanches of activity\citep{Bingham2021}.
An in-plane field will advance ASI state primarily when the strength of the field is increased beyond the coercivity of the array, and is highly dependent on field resolution\citep{Ke2008,Perrin2016,Li2018}.
Consequently, the discrete spin flip dynamics in the ASI are sudden and hard to control.

Here, we introduce a new field protocol scheme called \emph{astroid clocking}, which produces fundamentally different spin flip dynamics.
Astroid clocking of an ASI results in a step-wise, gradual evolution of spin states.
This offers unprecedented control and understanding of the dynamical process in both time and space.
By exploiting the shape and orientation of the nanomagnet switching astroids and their dipolar coupling, specific field angles are employed to selectively address different parts of the ensemble.
A clock protocol pulses fields at these angles in an alternate fashion, driving the intrinsic dynamics of the ASI.
Distinctively, the clock pulses maintain a constant field amplitude.

In the context of nanomagnetic logic, \citet{Nomura2017} demonstrated how the shape of two overlapping Stoner-Wohlfarth astroids can be exploited to preferentially switch nanomagnets in a 1D shift register.
Astroid clocking extends and generalizes this concept to 2D nanomagnet arrays and non-elliptical nanomagnets with different astroid shapes.
We show how astroid clocking reveals the intrinsic dynamics of coupled nanomagnetic systems.

For this study we consider the pinwheel ASI system, but stress that astroid clocking is readily applicable to other coupled nanomagnetic systems as well.
We demonstrate and analyse how ferromagnetic domains in pinwheel ASI can be gradually grown and reversed at will using astroid clocking.
Different clock protocols are explored, giving rise to distinct properties of the spin flip dynamics.

\section{Astroid clocking}

\begin{figure*}
    \includegraphics[width=\textwidth]{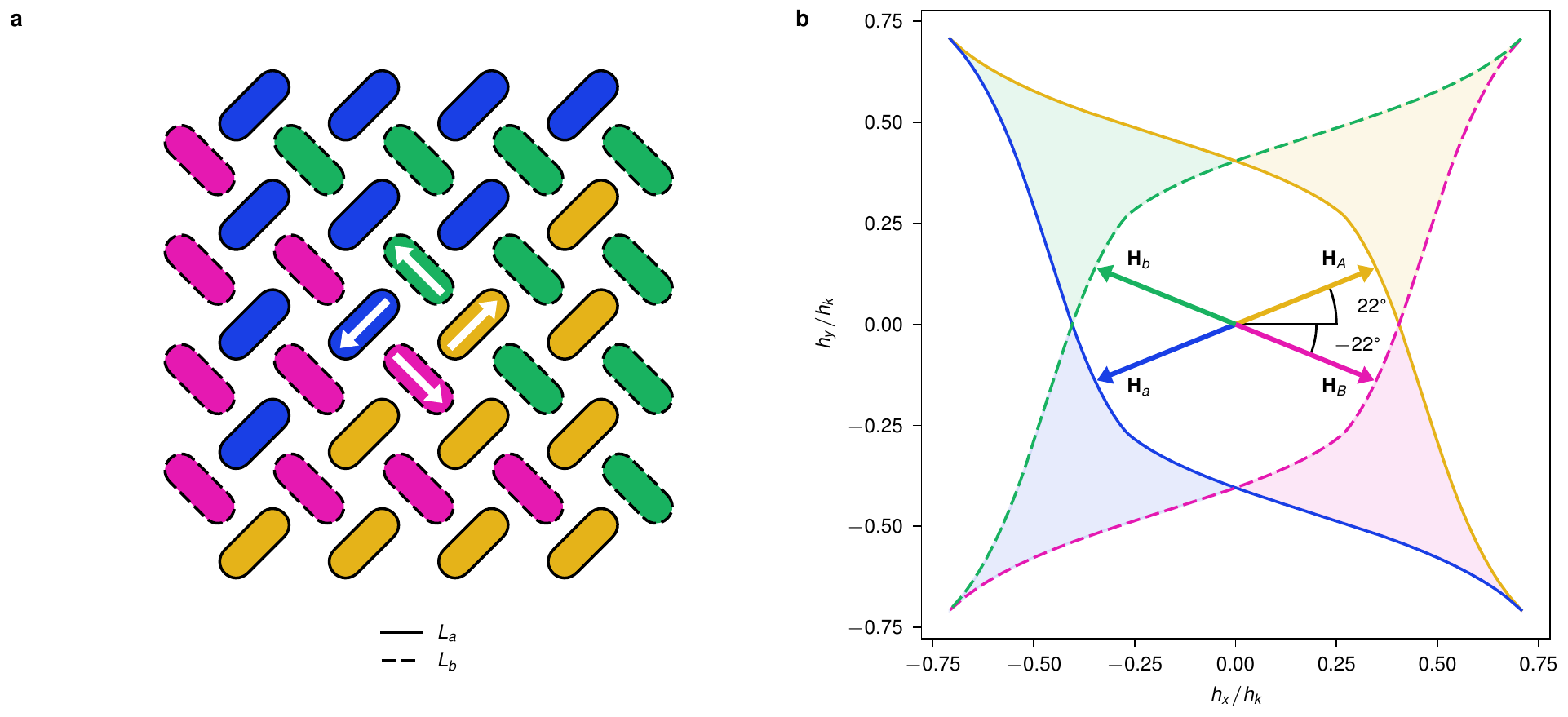}
    \caption{\label{fig:astroid-clocking}%
    Astroid clocking of pinwheel ASI.
    \textbf{a}, a small $4\times 4$ pinwheel ASI, formed by two interleaved sublattices $L_a$ (solid outline) and $L_b$ (dashed outline) with magnets oriented at $+45\degree$ and $-45\degree$, respectively.
    Colors correspond to magnetization direction as indicated by the white center arrows.
    The magnetic state shows the four possible ferromagnetic domains of pinwheel ASI, where the net magnetization forms a counter-clockwise magnetic flux closure pattern.
    \textbf{b}, switching astroids of the magnets in sublattice $L_a$ (solid lines) and $L_b$ (dashed lines), along with the four clock fields, $\vec{H}_A$, $\vec{H}_B$, $\vec{H}_a$, and $\vec{H}_b$.
    The astroid edges are colored according to the magnet state which is promoted when fields cross the edge.
    Similarly, the colored regions correspond to fields that \emph{exclusively} promote a magnet state within a sublattice.
    Astroid axes are normalized with respect to the hard axis switching threshold, $h_k$.
    }
\end{figure*}

Pinwheel ASI\citep{Gliga2017,Macedo2018,Li2018} consists of nanomagnets arranged on two interleaved square sublattices, as shown in \cref{fig:astroid-clocking}a.
In this study, the magnets in the two sublattices $L_a$ and $L_b$ are rotated $+45\degree$ and $-45\degree$ with respect to the array edges.
The sublattice and magnetization of the magnets are indicated by their color: magnets in sublattice $L_a$ are orange or blue, while magnets in sublattice $L_b$ are pink or green.
For brevity, we will refer to magnet state by these four colors.

Pinwheel ASI favors a ferromagnetic ordering, with emergent domains of coherent magnetization.
\cref{fig:astroid-clocking}a shows the four possible domain directions: rightwards (orange/pink), leftwards (blue/green) and so on.
The ferromagnetic domains are separated by domain walls, which are slightly less energetically favorable\citep{Macedo2018}.

The switching threshold of a nanomagnet depends on the field angle, and can be approximated by the Stoner-Wohlfarth astroid.
\cref{fig:astroid-clocking}b shows the switching astroids for the two orientations of stadium-shaped magnets in pinwheel ASI\citep{Flatspin2022}.
A magnet will switch state if the total field acting on it lies outside the astroid boundary, \emph{and} the field is directed against the current magnetization.

Nanomagnet shape largely determines the shape of the astroid.
Stadium-shaped nanomagnets, commonly used in ASI, have a switching astroid with 2-fold rotational symmetry\citep{Flatspin2022}.
This is in contrast to classical Stoner-Wohlfarth astroids that display 4-fold rotational symmetry derived for elliptical nanomagnets\citep{Tannous2008}.

Switching astroids that break the 4-fold rotational symmetry, can be exploited to selectively address nanomagnets that are rotated $90\degree$ with respect to each other.
If the total field lies within the shaded regions in \cref{fig:astroid-clocking}b, \emph{only} the nanomagnets in the corresponding sublattice will be able to switch.
A field in the orange/blue shaded regions will address only the magnets in sublattice $L_a$, while a field in the pink/green regions will address only magnets in $L_b$.
Furthermore, each region promotes a specific magnet state within each sublattice, e.g., a field in the blue shaded region promotes blue magnets by switching orange magnets.

In this study, we define two \emph{bipolar clocks} A and B along the $+22\degree$ and $-22\degree$ axes, respectively.
As shown in \cref{fig:astroid-clocking}b, each clock consists of a positive and negative \emph{clock field} of magnitude $H$ along the clock axis.
The four arrows in \cref{fig:astroid-clocking}b are colored according to the magnet states they promote, e.g., the $\vec{H}_A$ field only promotes orange magnets.

The dipolar fields $\vec{h}_\text{dip}$ from neighboring magnets may either promote or prevent switching.
If the dipolar fields are directed out of (into) the astroid, they effectively promote (prevent) switching.
A clock field can thus selectively address a subset of a sublattice, depending on the state of the ensemble.
The clock angles $\pm22\degree$ are selected to allow the dipolar fields to have a large influence on switching, using a field strength $H$ close to the switching threshold.
However, a precise angle is not crucial and the method tolerates a wide range of clock angles. 
Our system tolerates clock angles in the range \SIrange{10}{35}{\degree}, and field strengths accurate to \SIrange{3}{4}{\milli\tesla}.

\begin{figure*}
    \centering
    \includegraphics[width=.75\textwidth]{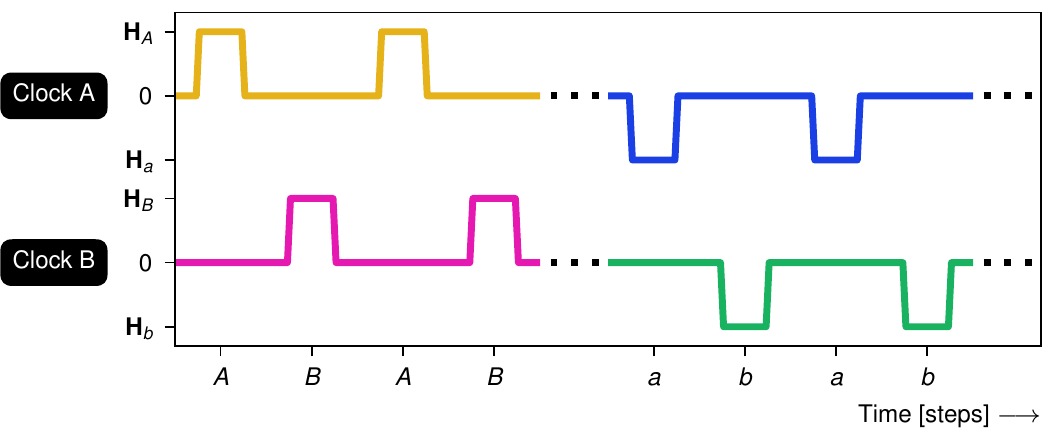}
    \caption{\label{fig:unipolar-clocking}%
    Clock diagram of astroid clocking.
    Clock protocols are defined by sequences of clock pulses.
    The clock diagram shows $AB$ clocking (alternating pulses of the positive clock fields $\vec{H}_A$ and $\vec{H}_B$) followed by $ab$ clocking (alternating pulses of the negative clock fields $\vec{H}_a$ and $\vec{H}_b$).
    }
\end{figure*}

\cref{fig:unipolar-clocking} illustrates astroid clocking, where a \emph{clock pulse} is defined as ramping a clock field from zero to $H$ and down to zero again.
The ramping speed is much slower than the timescale of nanomagnetic switching.
A \emph{clock protocol} is a specific sequence of clock pulses.
For example, $AB$ clocking consists of repeated alternating clock pulses of $A$ and $B$.
We define a \emph{clock cycle} as a single sequence of the clock pulses in a protocol, e.g., an $aAbB$ clock cycle is the sequence of four pulses $(a, A, b, B)$.
A \emph{unipolar} clock protocol exclusively employs one polarity of each clock, while a \emph{bipolar} clock protocol employs both polarities.

\section{\label{sec:unipolar-clocking}Unipolar clocking}

\begin{figure*}
    \includegraphics[width=\textwidth]{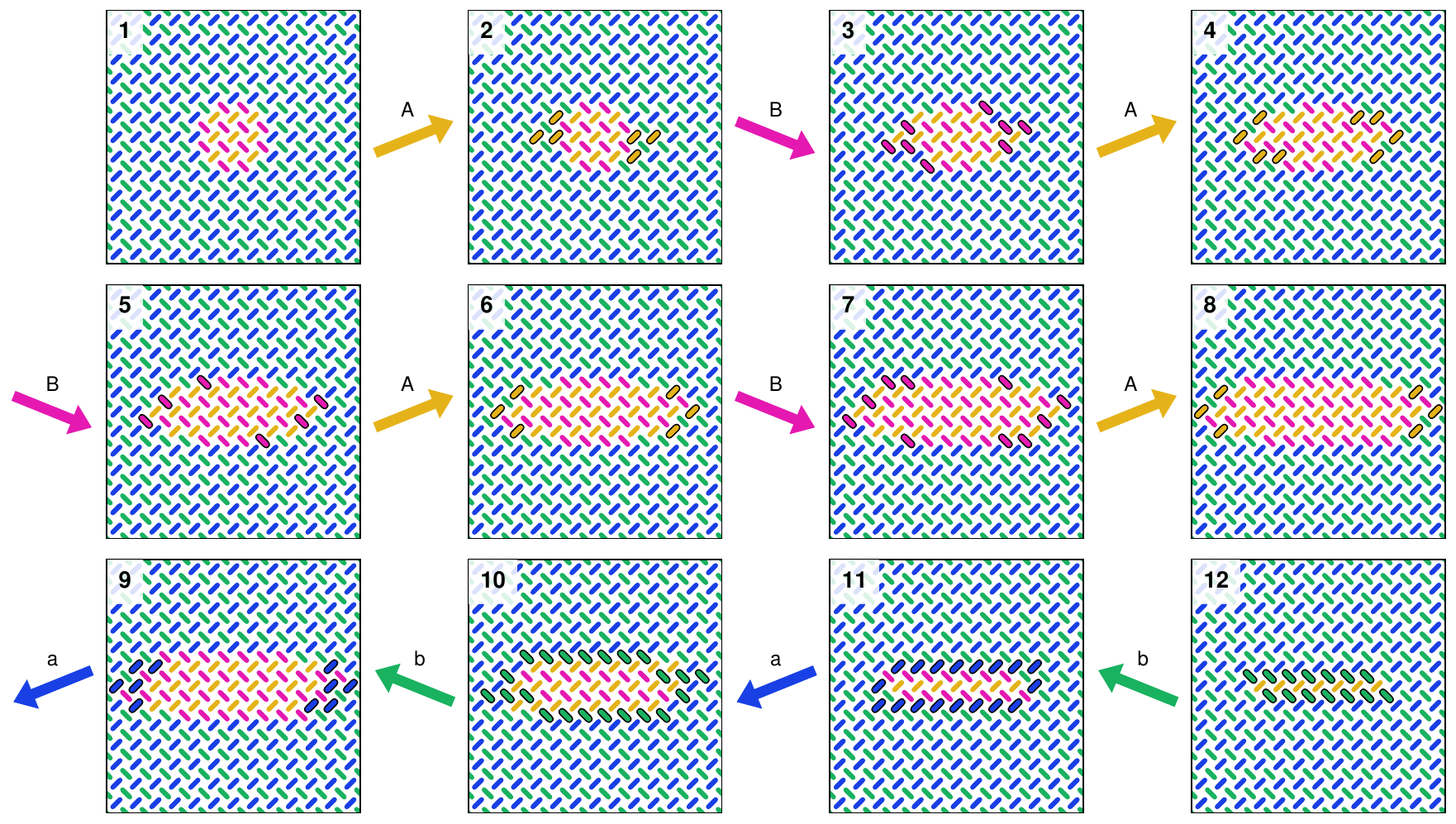}
    \caption{\label{fig:AB-clocking}%
    Simulation of unipolar astroid clocking of pinwheel ASI in flatspin. 
    Each snapshot shows a zoomed-in view of the $50\times50$ nanomagnet system, at different points during a clock protocol.
    (1) shows the initial state, a small orange/pink (rightwards) domain in the center of an otherwise polarized blue/green (leftwards) array.
    (2-8) show the state during $AB$ clocking, resulting in gradual domain growth.
    (9-12) show the subsequent states during $ab$ clocking, resulting in gradual domain reversal.
    Magnets that change state between snapshots are highlighted with a solid black outline.
    }
\end{figure*}

First, we explore the spin flip dynamics of pinwheel ASI when subject to the unipolar clock protocols $AB$ and $ab$.
The $50\times 50$ pinwheel ASI (5100 magnets) is initialized with a small rightwards (orange/pink) domain in the center of an otherwise leftwards polarized (blue/green) array.
\cref{fig:AB-clocking} (1) shows a closeup of the initial state.

\cref{fig:AB-clocking} (2-8) shows the state evolution of the array subject to $AB$ clocking, obtained from flatspin simulations (see \cref{sec:methods-flatspin}).
As expected, the $A$ pulse selectively switches magnets in sublattice $L_a$ from blue to orange, while the $B$ pulse selectively switches magnets in sublattice $L_b$ from green to pink.

Interestingly, the particular magnets that switch are the ones along the domain wall.
As a result, the inner (leftwards) domain grows gradually over time, with only a thin layer of the domain advancing after each clock pulse.
The growth is \emph{monotonic} and \emph{step-wise}, driven by the clock pulses.

A curious property is that the domain grows mainly in the horizontal direction.
In \cref{fig:AB-clocking} (2-8), the magnets along the vertical domain walls are the only ones to switch.
If the growing domain reaches the edges of the array, the direction of growth changes and becomes vertical, eventually filling the entire array (\cref{fig:AB-clocking-edge}).

Inverting the clock pulses ($ab$ clocking), will instead grow the outer (blue/green) domain and consequently shrink (reverse) the inner (orange/pink) domain.
As can be seen in \cref{fig:AB-clocking} (9-12),
domain reversal from (8) proceeds in both vertical and horizontal directions, resulting in reversal of the inner domain in fewer clock cycles compared to growth.
Hence there is an apparent \emph{asymmetry} in the direction of domain growth and reversal.

\section{\label{sec:mechanism}Growth and reversal mechanism}

\begin{figure*}
    \includegraphics[width=\textwidth]{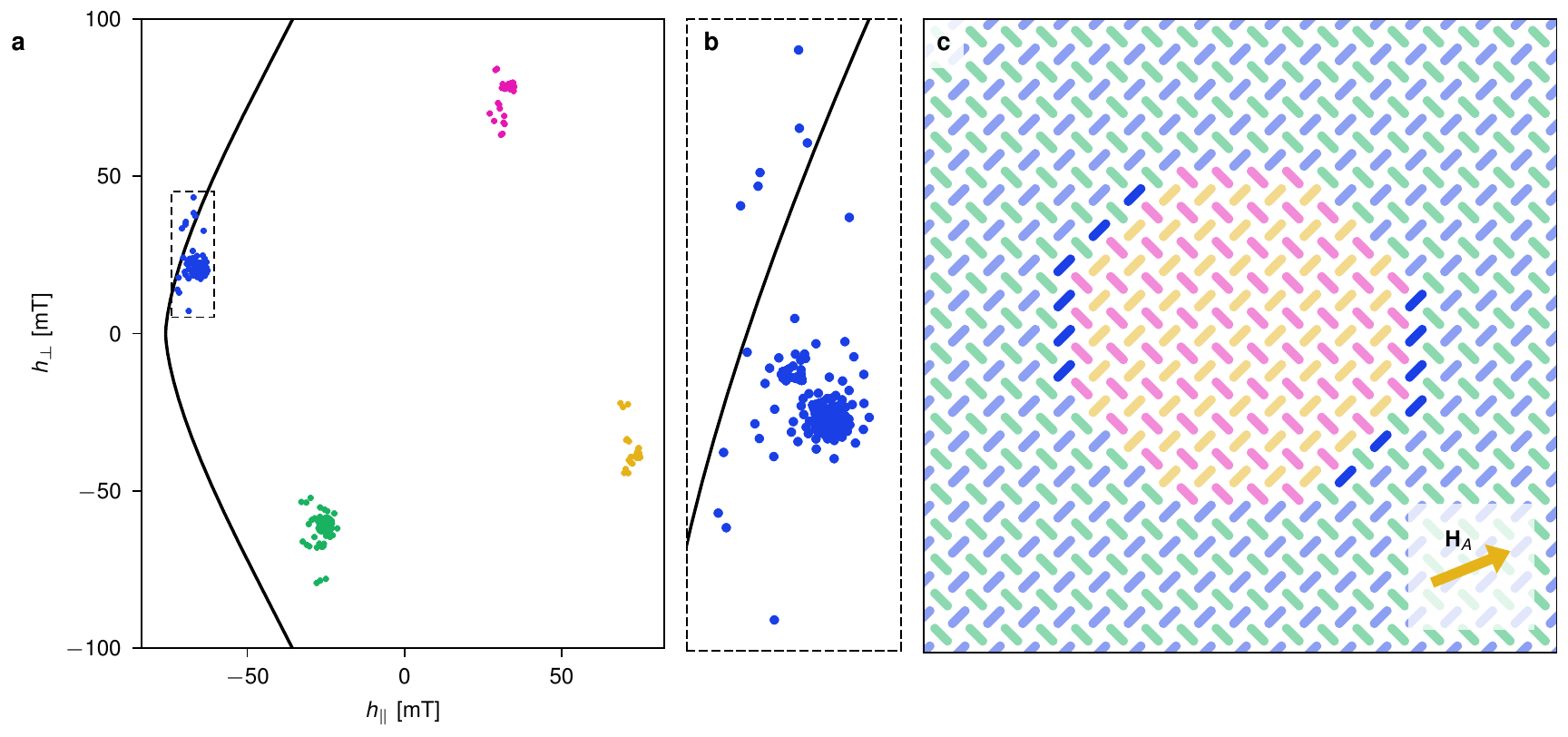}
    \caption{\label{fig:astroid-clusters}%
    Astroid clusters and astroid (black curve) for the pinwheel system shown in \textbf{c}, when subject to the clock field $\vec{H}_{A}$.
    \textbf{a}, astroid cluster plot where each dot represents the total field $\vec{h}_i = \vec{H}_{A} + \vec{h}_\text{dip}^{(i)}$ experienced by a magnet $i$, projected onto its parallel ($h_\parallel$) and perpendicular ($h_\perp$) axis.
    The colors in the plot correspond to magnet state.
    Note that the astroid plot shows location \textit{relative} to each magnet's own switching threshold, e.g., orange magnets are far from switching as they are aligned with $\vec{H}_A$.
    \textbf{b}, closeup of the blue cluster, revealing a sub-group of blue magnets that lie outside the switching astroid and are eligible for switching.
    These magnets are highlighted in \textbf{c}, and are all found to lie along the vertical and $+45\degree$ domain walls.
    The data is obtained from flatspin simulations.
    }
\end{figure*}

To understand the mechanism behind the domain growth and reversal, we consider the larger domain shown in \cref{fig:astroid-clusters}c, subject to $\vec{H}_A$.
In \cref{fig:astroid-clusters}a, we plot the relative locations of all the magnets within their respective switching astroids.
Each dot represents the total field $\vec{h}_i = \vec{H}_A + \vec{h}_\text{dip}^{(i)}$ experienced by a magnet $i$ in its \emph{local frame of reference}.
There are four clusters of dots within the astroids, corresponding to the four magnet colors, where only the blue magnets are close to switching.

The internal structure of each astroid cluster is a result of the nanomagnet dipolar coupling, and a direct consequence of the ASI geometry.
In the absence of dipolar fields, each cluster collapses into a single point.
The dipolar fields add complex structure to the clusters, with sub-groups corresponding to different subsets of magnets within the ASI.
For a detailed analysis of neighbor contributions, see \cref{sec:supplementary-neighbor-interactions}.

The inset shown in \cref{fig:astroid-clusters}b reveals the structure of the blue cluster.
Notice there are a few blue dots that lie outside the astroid, corresponding to magnets that are eligible for switching, which are highlighted in \cref{fig:astroid-clusters}c.
Evidently, the switchable magnets all lie along the vertical and $+45\degree$ domain walls.

When a magnet switches, its location within the astroid jumps to the cluster of opposite spin, e.g., a blue magnet switches to the orange state.
In addition, neighboring magnets will see a change in the dipolar fields, causing movement within their respective clusters.
In this way, the switching of a magnet may enable future switching in neighboring magnets, either during the current or a future clock pulse.

The observed horizontal domain growth can now be explained from the internal structure of the astroid clusters.
We have seen that magnets along certain types of domain walls can be selectively switched under an applied clock field.
Switching the blue (highlighted in \cref{fig:astroid-clusters}c) magnets along these domain walls reverses their dipolar fields, which affects the structure of the green cluster.
Consequently, green magnets that are part of the domain walls will approach the switching astroid.
When the $\vec{H}_B$ field is subsequently applied, these magnets will be outside the astroid and hence switch.
As this cycle repeats, the result is an apparent horizontal domain growth emerging from the dipolar interactions and clock fields.

\begin{figure}[h]
    \includegraphics[width=\textwidth]{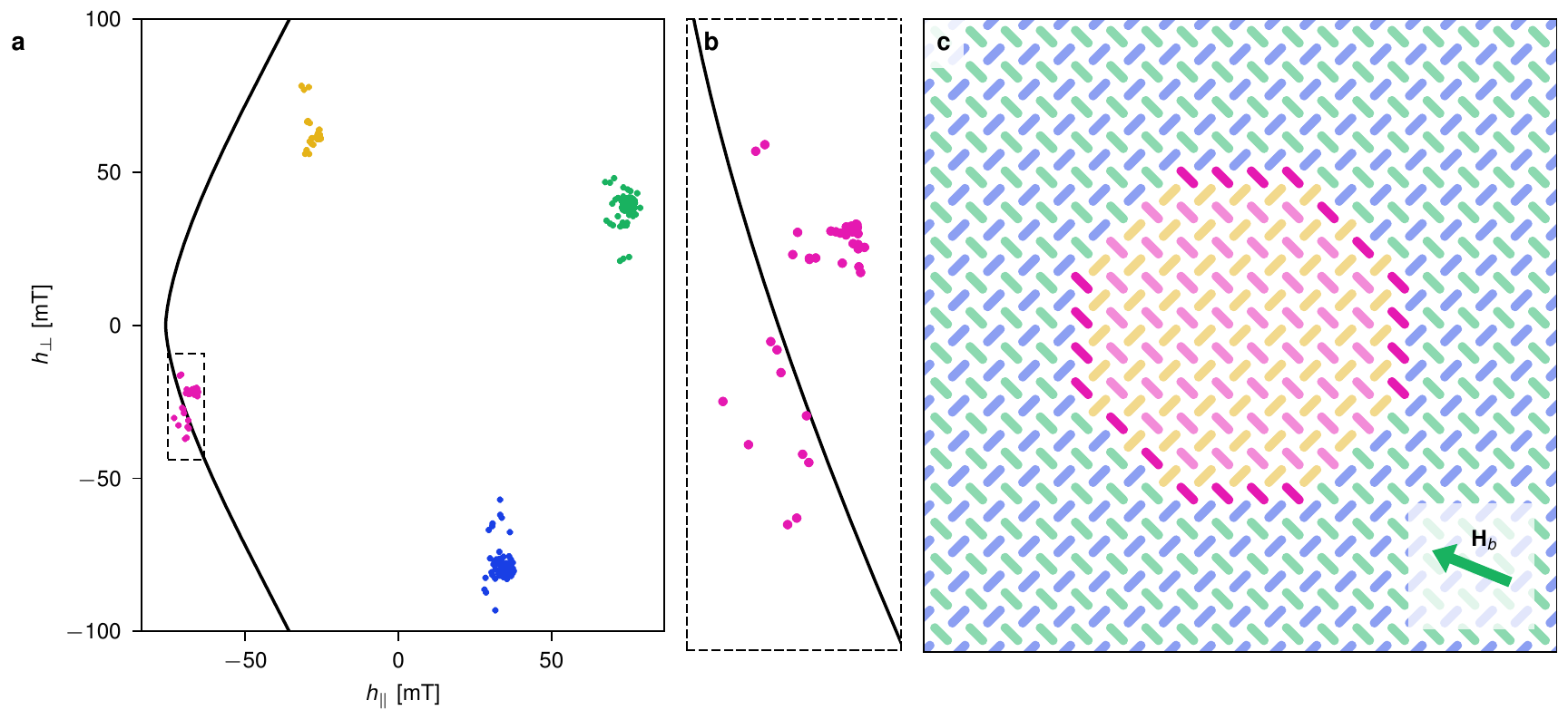}
    \caption{\label{fig:astroid-clusters-reversal}%
    \textbf{a-b}, astroid clusters during reversal, when the pinwheel system shown in \textbf{c} is subject to the negative clock field $\vec{H}_b$.
    \textbf{b}, astroid clusters during reversal have a different structure compared to growth.
    Switchable magnets outside the astroid are highlighted in \textbf{c}.
    During reversal, the switchable magnets are along both the horizontal, vertical and $-45\degree$ domain walls.
    Switchable magnets along the horizontal domain wall is attributed to the curvature of the inner domain.
    }
\end{figure}

During domain \textit{reversal}, both horizontal and vertical domain walls take part in the process.
As a result, reversal requires fewer clock cycles compared to growth.
During reversal, the switchable magnets lie along both the horizontal, vertical and $-45\degree$ domain walls (\cref{fig:astroid-clusters-reversal}).

We find that the horizontal domain wall movement, particular to reversal, is dependent on the curvature of the reversing domain.
If the horizontal domain wall is surrounded by a blue/green domain on three sides, there is a stronger dipolar ``push'' towards the astroid edge.
As such, domain shape plays a crucial role in the reversal process.

\begin{figure}[h]
    \includegraphics[width=\textwidth]{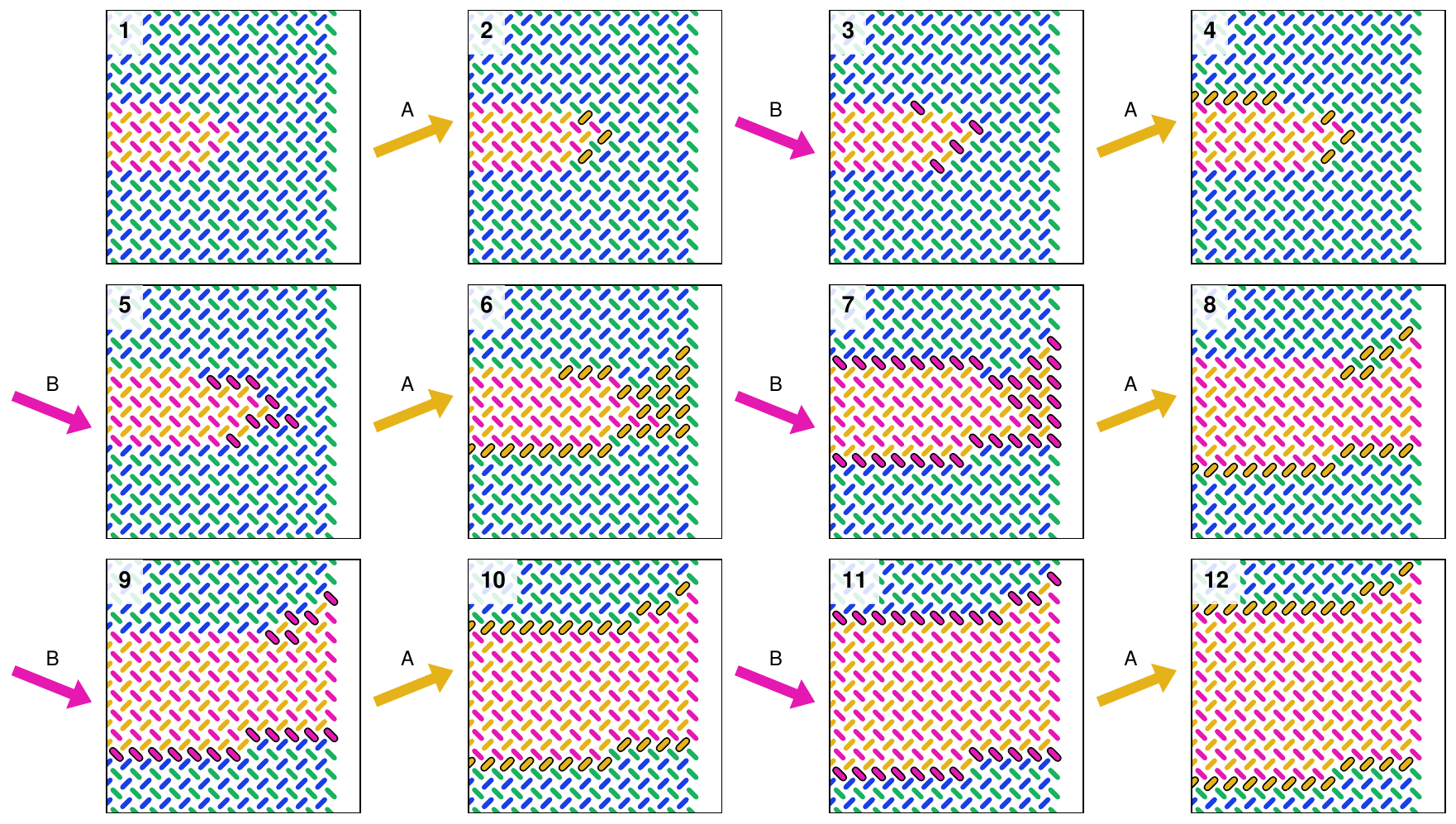}
    \caption{\label{fig:AB-clocking-edge}%
    Unipolar $AB$ clocking of an orange/pink (rightwards) domain as it reaches the edge of the array.
    There is an apparent transition from horizontal to vertical domain growth (5-6).
    Vertical growth proceeds by avalanches of spin flips, starting at the bottom-left and top-right corners of the domain at the array edge.
    Magnets that change state between snapshots are highlighted by a solid black outline.
    }
\end{figure}

When a domain grows to reach the edges of the array, there is an apparent transition from horizontal to vertical growth (\cref{fig:AB-clocking-edge}).
We find that vertical growth proceeds by avalanches along the domain wall, starting at the bottom-left and top-right corners of the domain, close to the array edges.

\section{\label{sec:exp-growth-reversal-results}Experimental growth and reversal}
Next, we demonstrate astroid clocking of pinwheel ASI experimentally.
Samples are imaged with x-ray magnetic circular dichroism photoemission electron microscopy (XMCD-PEEM), with an in-situ vector magnet to perform astroid clocking. 
See \cref{sec:methods} for details.

\begin{figure*}
    \centering
    \includegraphics[width=1.0\textwidth]{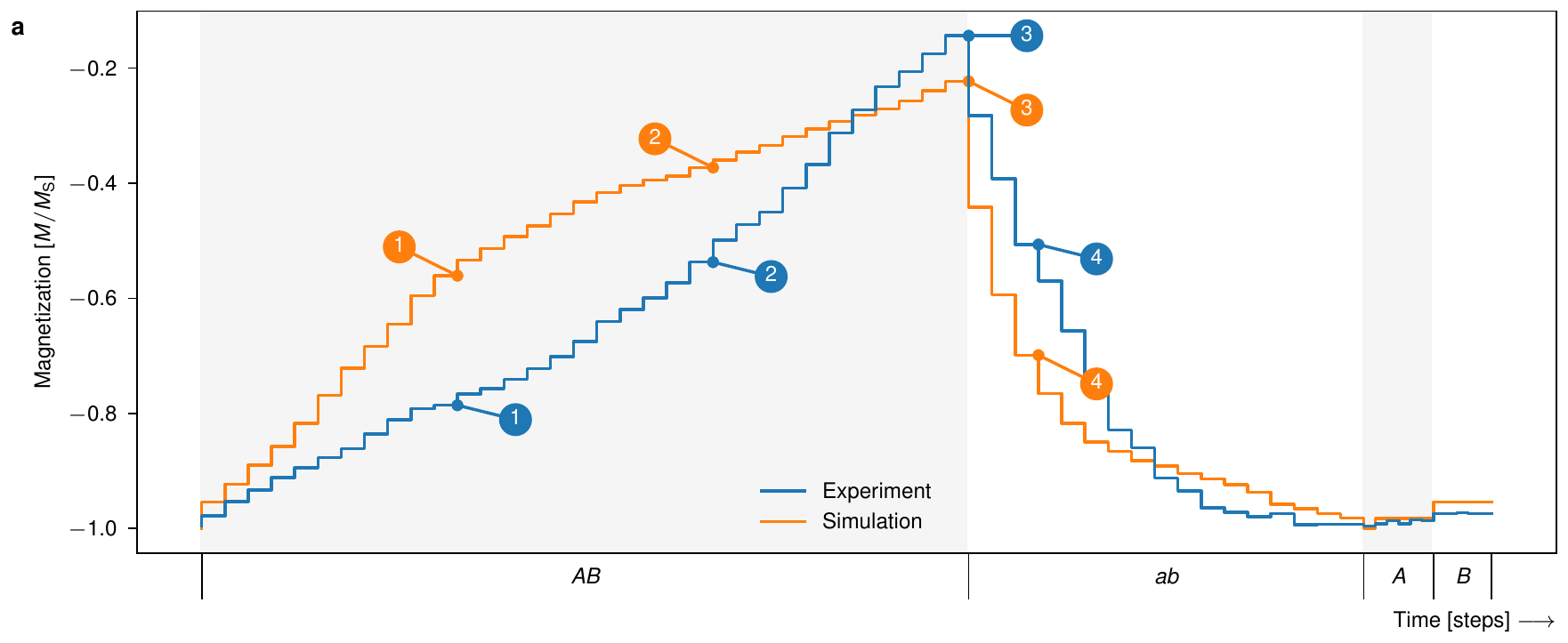}
    \includegraphics[width=1.0\textwidth]{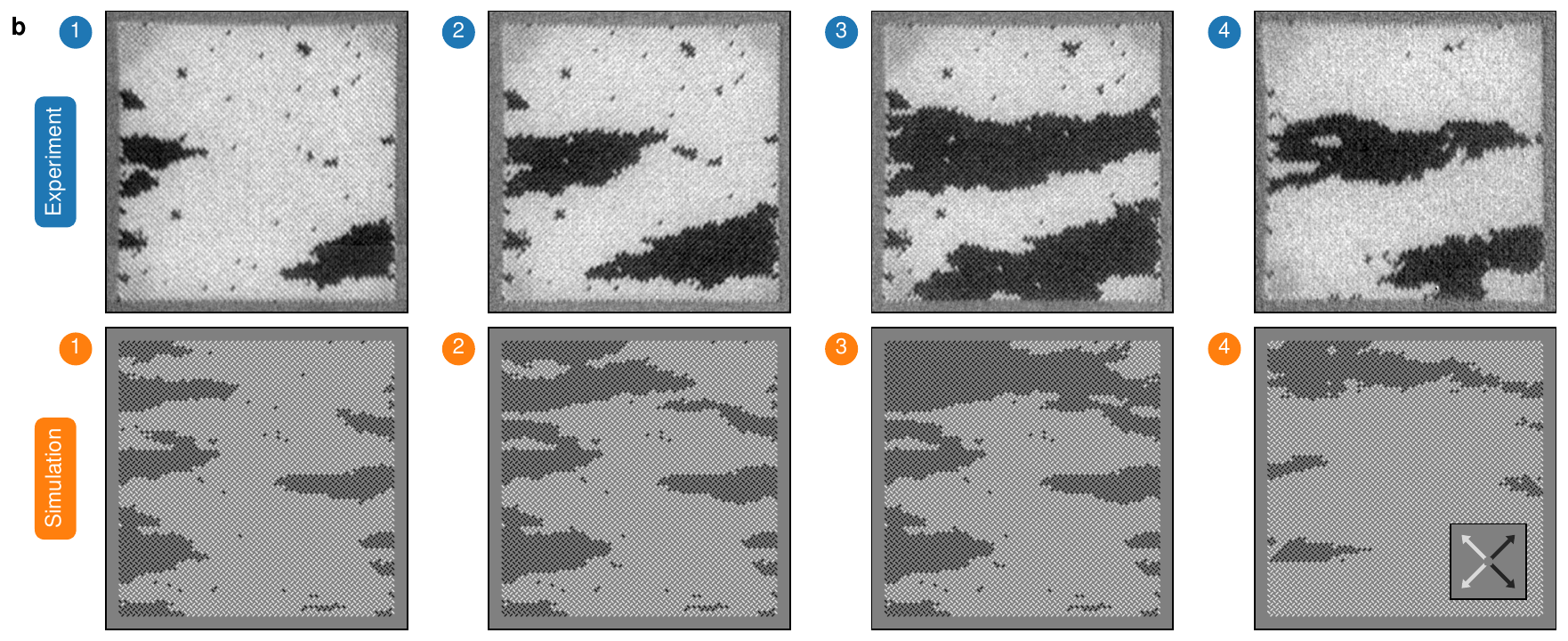}
    \caption{\label{fig:ab-series}%
    Results of growth and reversal with unipolar clock protocols, and control experiment. 
    \textbf{a}, total magnetization of the ensembles subject to the different clock protocols.
    The timeline indicates clock time, labeled by the clock protocol. 
    During $AB$ clocking, the ensembles undergo growth and hence an increase in magnetization.
    The second phase, $ab$ clocking, quickly reverses domains and total magnetization.
    The control experiment, consisting of separate $A$ and $B$ clock sequences, show no development of the domains.
    \textbf{b}, magnetic image snapshots (experiemental XMCD-PEEM images and flatspin simulated XMCD-PEEM contrast images) of the ensembles at the specified points in time. 
    The depicted ensembles are approximately \qtyproduct{12.5x12.5}{\micro\meter} ($50\times50$ pinwheel ASI, 5100 magnets).
    All XMCD-PEEM images are available in \cref{fig:all-AB-snapshots-1,fig:AB-control-snapshots}.
    Videos of the experiment and simulation are provided in \cref{vid:unipolar-peem,vid:unipolar-sim}. 
    }
\end{figure*}

After polarizing all magnets in the leftwards direction (bright contrast), we perform steps of $AB$ clocking, imaging in-between each clock cycle.
\cref{fig:ab-series}a shows total magnetization of the ensemble, obtained from the XMCD-PEEM images, which increases in a stable, monotonic fashion.
Selected experiment snapshots are shown in \cref{fig:ab-series}b.
Snapshots (1-3) show that domains nucleate at the vertical edges then predominantly expand horizontally.

Domain formation at the \textit{vertical} array edges can be explained by the dipolar field-driven mechanism behind $AB$ clocking.
While domain nucleation along horizontal edges is possible, continued growth primarily occurs in the horizontal direction, preventing further expansion of horizontal edge nucleated domains.

In any physical ASI system, the nanomagnets will exhibit a range of intrinsic switching thresholds, a \textit{disorder}, due to imperfections and microscopic variations of material composition. 
Disorder affects both domain shape and growth dynamics, as evident in our experimental results.
Compared to the idealized simulations, domains appear more organic, with distinct features such as jagged edges, slanted domain walls, and sporadic holes.
In terms of dynamics, some domain borders get stuck for several clock cycles, while others advance more than one step during a single cycle (see \cref{fig:all-AB-snapshots-1}).

By introducing disorder to the simulations (see \cref{sec:methods-flatspin}), we obtain results that more closely resemble the experiment.
The magnetization curve and snapshots from simulations \textit{with disorder} are included in \cref{fig:ab-series}.
Notice how the simulated snapshots show organic-looking domains that resemble the domains of the experiment.

After growth, we apply the reversal clock protocol, $ab$ clocking.
For each $ab$ clock cycle, the magnetization reduces sharply, with domains shrinking more rapidly compared to the increase during growth. 
Comparing snapshot 3 and 4 of \cref{fig:ab-series}b, it is clear that the domains shrink in both vertical and horizontal directions.

Next, we conduct a control experiment to verify that simply repeating a clock pulse $A$ or $B$ does not result in domain growth.
After re-initializing the system, we apply several pulses of $A$, then several pulses of $B$, imaging after each pulse.
As seen in \cref{fig:AB-control-snapshots} and the last part of \cref{fig:ab-series}a, only the first application of $A$ or $B$ results in growth.
Growth progresses only when the type of clock pulse is changed, which confirms that the alternating pattern of $A$ and $B$ is what drives the observed domain growth.

These experiments affirm the viability of astroid clocking in the face of experimental sensitivities (as low as \SI{<1}{\milli\tesla} from \cref{fig:astroid-clusters}) and potential impediments such as fabrication imperfections, temperature effects, and material degradation.
While unstable individual magnets and inaccuracies in the image analysis induce some noise, it is negligible compared to the effect of astroid clocking.
Experimental astroid clocking is surprisingly robust, demonstrating that it is possible to precisely control the spin flip dynamics of ASIs using global fields.

\section{\label{sec:bipolar-clocking}Bipolar clocking}

\begin{figure*}
    \centering
    \includegraphics[width=.75\textwidth]{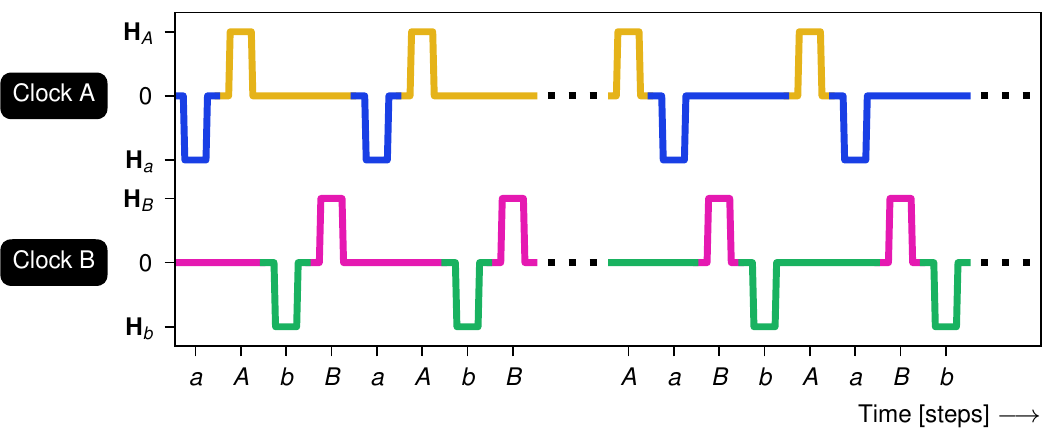}%
    \caption{\label{fig:bipolar-clocking}%
    Clock diagram of bipolar $aAbB$ clocking followed by its inverse, $AaBb$ clocking.
    Bipolar clocking employs both positive and negative clock pulses.
    }
\end{figure*}

In bipolar clocking, each clock may be pulsed in both polarities.
We consider two clock protocols illustrated in \cref{fig:bipolar-clocking}, namely $aAbB$ and its inverse, $AaBb$ clocking.
In contrast to unipolar clocking, the magnetic fields in these bipolar clock protocols are balanced, i.e., the sum of all clock fields is zero.
One might then expect that this results in a net zero magnetization change.

\begin{figure}[h]
    \centering
    \includegraphics[width=1.0\textwidth]{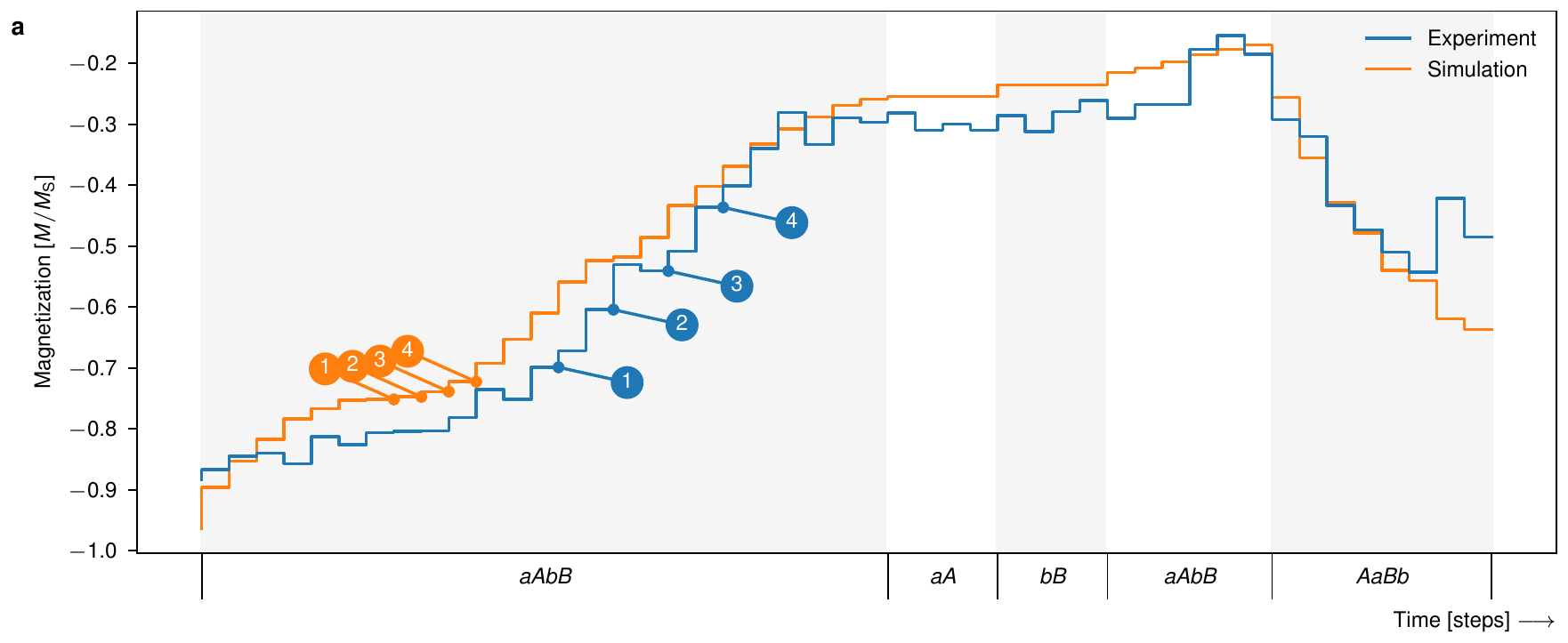}
    \includegraphics[width=1.0\textwidth]{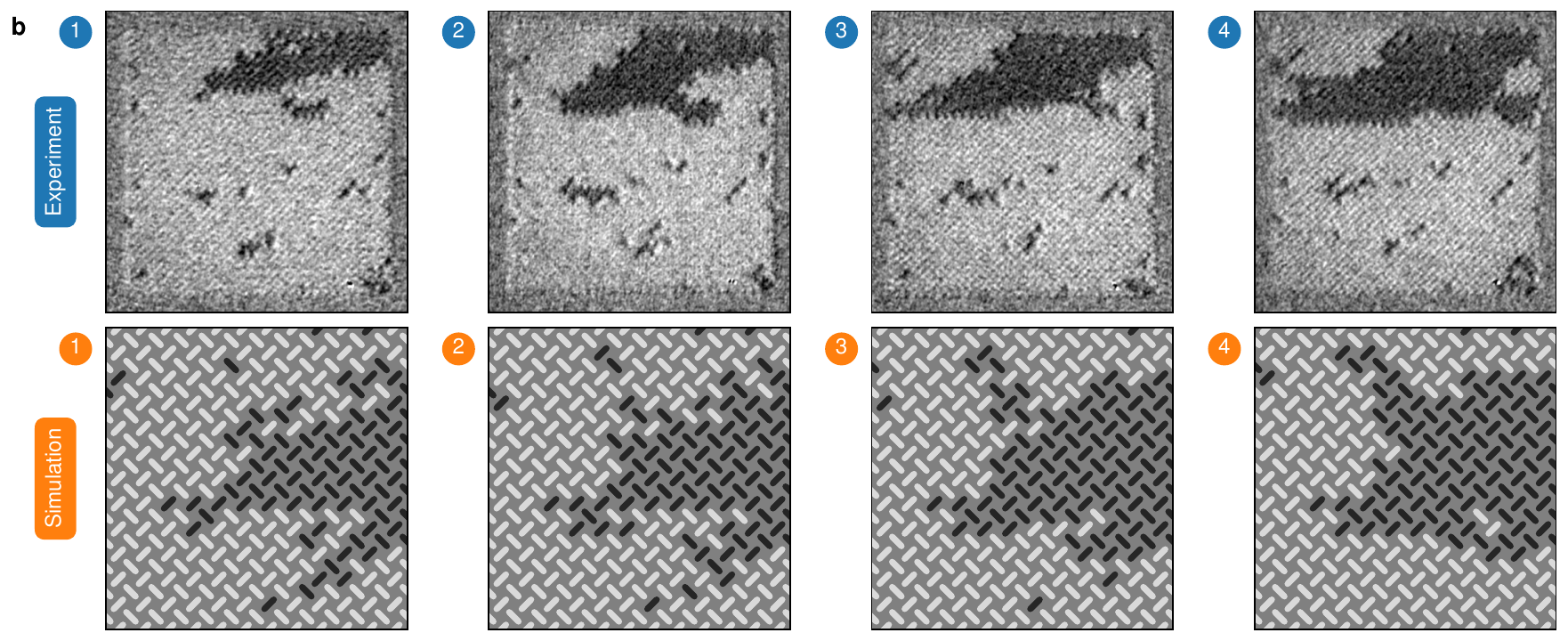}
    \caption{\label{fig:aAbB-series}%
    Results of growth and reversal with bipolar clocking, and control experiment.
    \textbf{a}, total magnetization of the ensembles subject to the different bipolar clock protocols.
    The timeline indicates clock time, labeled by the clock protocol. 
    During the first phase, $AaBb$ clocking, the ensembles undergo domain growth and increase in magnetization.
    The controls, $aA$ clocking and $bB$ clocking, show no net growth.
    Further growth ($aAbB$ clocking) and reversal ($AaBb$ clocking) occur after the controls. 
    \textbf{b}, magnetic image snapshots of the experimental ensemble, and zoomed in views of the flatspin simulated ensemble, at the specified points in time.
    The growing domains change morphology during the clock protocol. 
    All XMCD-PEEM images are available in \cref{fig:all-DACB-snapshots}.
    Videos of the experiment and simulation are provided in \cref{vid:bipolar-peem,vid:bipolar-sim}.
    }
\end{figure}

On the contrary, bipolar clocking also results in domain growth and reversal, and a net change in magnetization.
\cref{fig:aAbB-series}a plots the total magnetization of pinwheel ASI subject to bipolar clocking.
As can be seen, $aAbB$ clocking results in net domain growth, while $AaBb$ clocking results in domain reversal.

In contrast to unipolar clocking, bipolar clocking can also induce morphological changes to the growing domains.
As a result of the bipolarity of the clock pulses, domains are now able to both grow and shrink within the same clock cycle.
In the experiment snapshots of \cref{fig:aAbB-series}b, we observe growth from (1) to (2), followed by a clear change in domain morphology from (2) to (3), and further growth between (3) to (4).
In simulations, we can observe the step-wise details of simultaneous growth and morphology changes, as shown in the zoomed in snapshots.
Inverting the clock protocol ($AaBb$ clocking) results in domain reversal.

The deciding factor for growth or reversal is the polarity of the last clock pulse at the transition between the two clocks.
Each clock in $aAbB$ clocking, for example, ends on the positive polarity at the transition ($aA$ and $bB$), resulting in growth of the rightwards (orange/pink) domains.

Within a bipolar clock cycle, there is an apparent competition between growth and reversal.
Some domain wall configurations result in net domain growth (others in net reversal), in a ``one step back, two steps forward'' process (see \cref{sec:supplementary-bipolar-process}).
In this way, a domain may grow horizontally and reverse vertically, thereby gradually changing shape over time (see \cref{fig:aAbB-clocking}).
While the balance between growth and reversal can be delicate, there is a clear trend for the clock protocols explored here, namely growth for $aAbB$, and reversal for $AaBb$.

Compared to unipolar clocking, the dynamics in bipolar clocked pinwheel ASIs are more varied and complex.
While there is a gradual net domain growth, the activity can intermittently spike and linger, depending on the particular state of the ensemble (see \cref{vid:bipolar-peem,vid:bipolar-sim}).
Bipolar clocking hence unlocks a wide variety of complex dynamic behavior in pinwheel ASI, while at the same time offering considerable control by choice of clock protocol.

\section{Conclusions}
We have introduced astroid clocking, a scheme for field-driven evolution in nanomagnetic metamaterials.
The method exploits the shape and orientation of the nanomagnet switching astroids and dipolar coupling to selectively address subsets of the nanomagnets.
Pulsing specific fields in sequence results in clocked dynamics that are both gradual and discrete in time.
Considerable control of the dynamics is available through choice of clock protocol.

This work demonstrates how astroid clocking can be used to control the growth and reversal of ferromagnetic domains in pinwheel ASI.
In this system, unipolar clocking results in monotonic domain growth or reversal, while bipolar clocking adds more complex dynamics that include changes to domain morphology.

The principles of astroid clocking are not limited to pinwheel ASI, and are applicable to a range of coupled nanomagnetic systems.
Exploring the clocked dynamics of established and future nanomagnetic metamaterials is an exciting research direction.
The space of possible clock protocols remains vast.

Astroid clocking offers unprecedented control and understanding of ASI dynamics in both time and space.
The method enables new directions in ASI research and paves the way for novel devices based on nanomagnetic metamaterials.

\newpage
\section{\label{sec:extended-figs}Extended data figures}

\begin{figure}[p]
    \centering
    \includegraphics[width=1.0\textwidth]{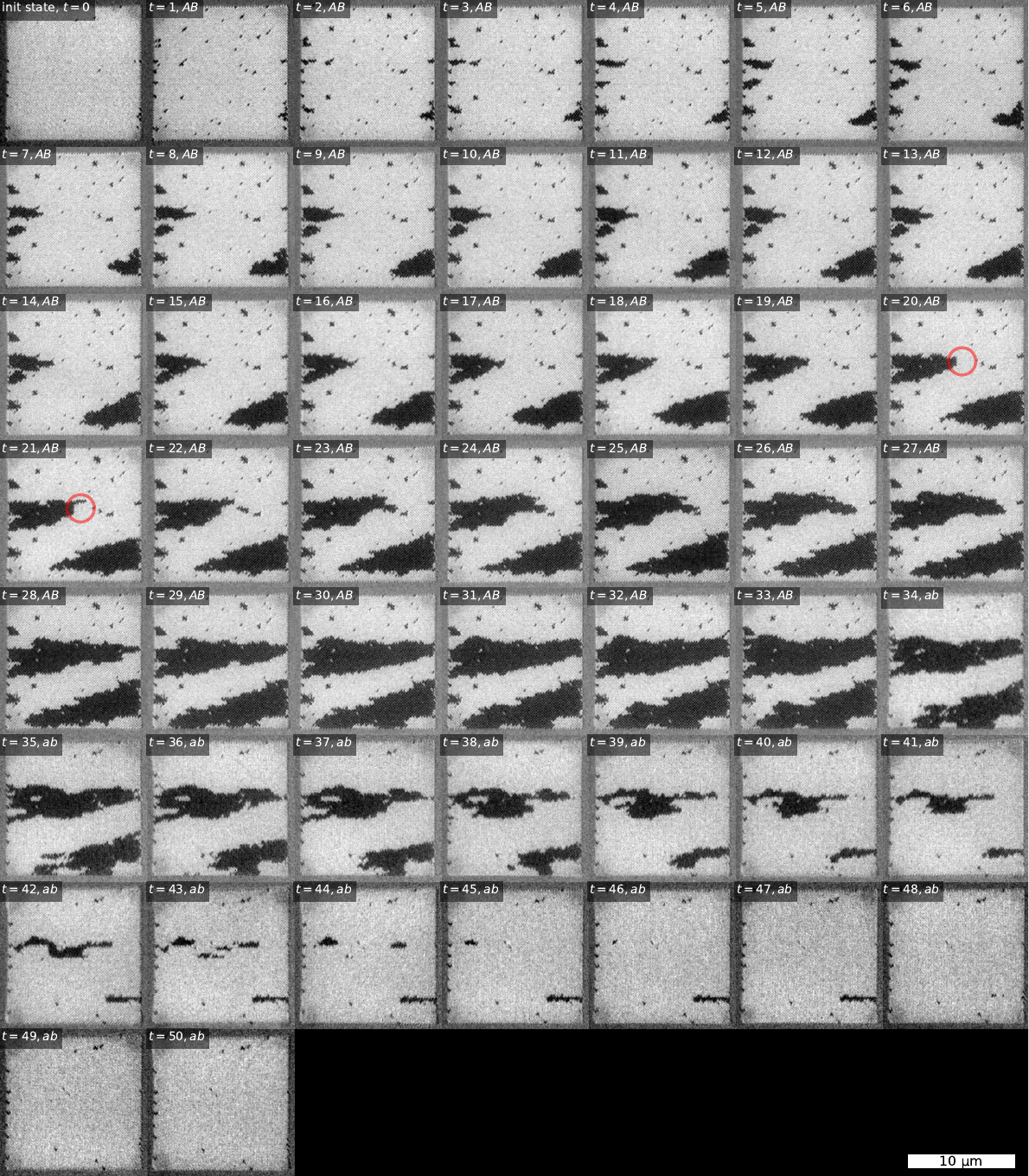}
    \caption{\label{fig:all-AB-snapshots-1}%
    XMCD-PEEM images of all steps from the relevant unipolar clock protocol series. 
    Time starts at $t=0$, and is incremented by 1 for each clock step, with clock pulses indicated by the labels. 
    The black (rightwards) domains grow with application of $AB$ clocking, and quickly reverses with $ab$ clocking.
    Red circle highlights: The short, vertical domain wall terminating the black domain in the center region of snapshot $t=20$ exemplifies both avalanching domain growth and a stuck domain wall. 
    In snapshot $t=21$ the top part of the domain wall has progressed in an avalanche to form a finger extension of the domain, while the bottom part of the domain wall remains as before.
    }
\end{figure}

\begin{figure}[p]
    \centering
    \includegraphics[width=1.0\textwidth]{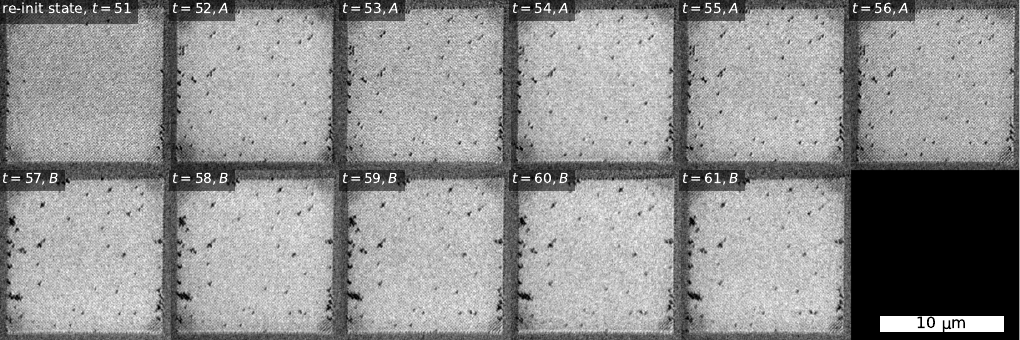}
    \caption{\label{fig:AB-control-snapshots}%
    XMCD-PEEM images of the control experiment. 
    The system is reinitialized at $t=51$ (following from \cref{fig:all-AB-snapshots-1}), and $t$ is incremented by 1 for each clock step, with clock pulses indicated by the labels. 
    The first $A$ clock pulse promotes dark (rightwards) magnets, equivalent to half a clock cycle, while subsequent applications of $A$ incurs no further change.
    When the clock pulse is changed to $B$, dark (rightwards) magnets are again promoted, equivalent to the second half of an $AB$ clock cycle. 
    Furthermore, additional $B$ clock pulses incurs no change in the state. 
    }
\end{figure}

\begin{figure}[h]
    \includegraphics[width=\textwidth]{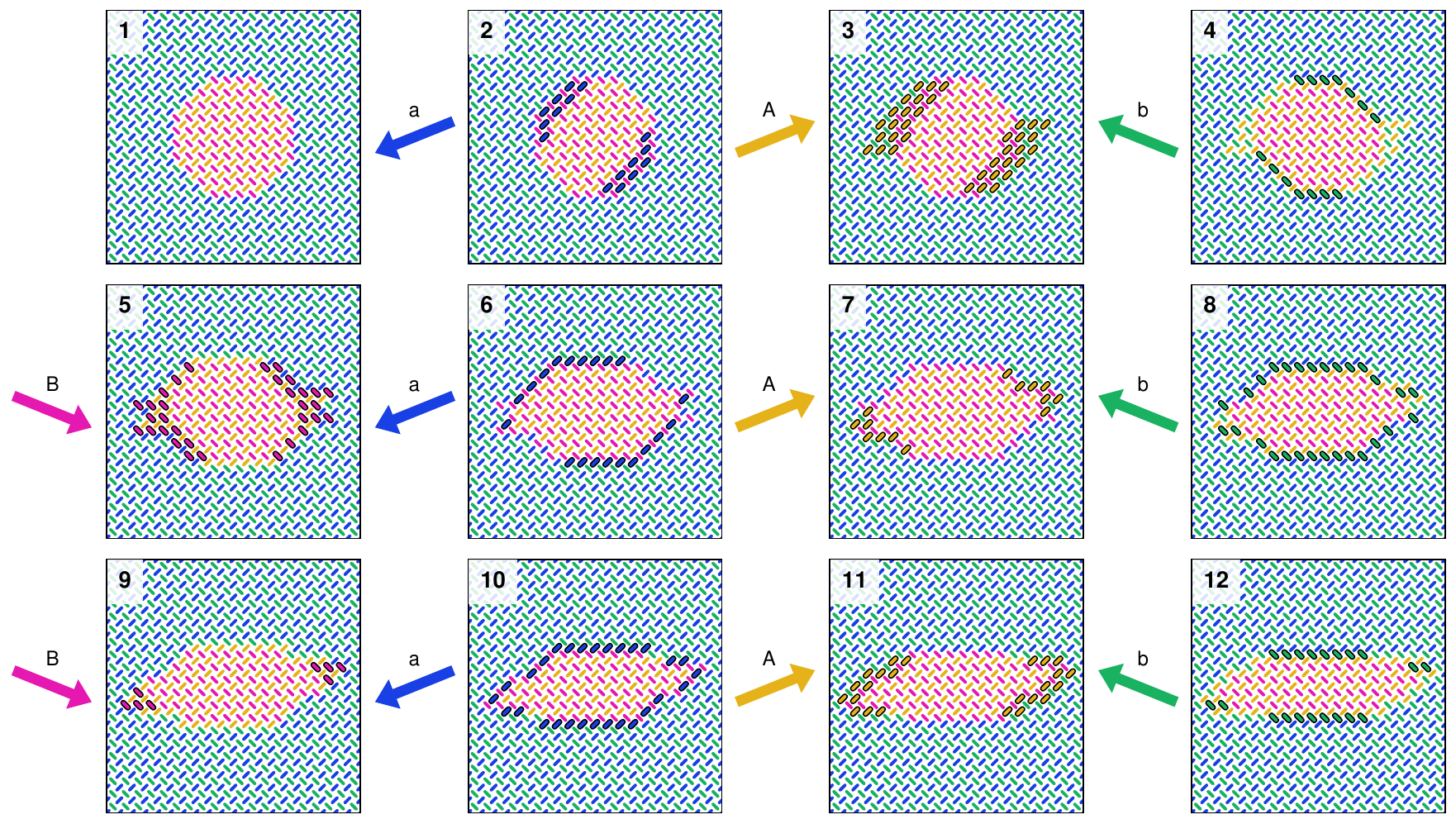}
    \caption{\label{fig:aAbB-clocking}%
    Bipolar $aAbB$ clocking of pinwheel ASI. 
    Each snapshot shows a zoomed-in view of a $50\times50$ system, at different points during a clock protocol.
    (1) shows the initial state, an orange/pink (rightwards) domain in the center of an otherwise polarized blue/green (leftwards) array.
    (2-12) show the state during $aAbB$ clocking, with simultaneous domain growth (horizontally) and reversal (vertically).
    As a result the domain gradually changes morphology over time.
    Magnets that change state between snapshots are highlighted by a solid black outline.
    }
\end{figure}

\clearpage

\section*{\label[methods]{sec:methods}Methods}
\subsection*{\label[methods]{sec:methods-sample-fabrication}Sample fabrication details}
The samples are arrays of permalloy nanomagnets fabricated in pinwheel ASI geometries on a silicon substrate. 
The resist mixture, 1:2 CSAR 62:anisole, is spin-coated onto the substrate at \SI{4000}{rpm}, achieving a thickness of \SI{\sim100}{\nano\meter}. 
Following coating, samples are soft baked at \SI{150}{\celsius} for 1 minute. 
The desired patterns, arrays of \qtyproduct{220 x 80}{\nano\meter} stadium shaped nanomagnets in $30 \times 30$ and $50 \times 50$ pinwheel geometries, are then exposed using the Elionix ELS-G100 EBL system.
Samples are post-exposure baked at \SI{150}{\celsius} for 1 minute.
The patterned resist is developed using AR600-546 for 1 minute, rinsed with isopropanol, and nitrogen dried. 
Permalloy (\ce{Ni_{0.79}Fe_{0.21}}) is deposited to a thickness of \SI{25}{\nano\meter} via electron beam evaporation using a Pfeiffer Vacuum Classic 500 system, and capped with a \SI{2}{\nano\meter} aluminium layer.
Finally, the samples undergo ultrasound-assisted lift-off using a dedicated stripper (AR600-71), leaving behind the patterned permalloy nanomagnets.
Post-fabrication, the precision and quality of the fabricated nanomagnet arrays are inspected using Scanning Electron Microscopy (SEM).
This SEM inspection confirmed that the permalloy nanomagnets are properly formed, free-standing, and without significant defects.

\subsection*{\label[methods]{sec:methods-clocking-procedure}XMCD-PEEM and clocking procedure}
Experimentally realized clocking of fabricated ASIs is carried out under magnetic microscopy inspection.
We use a photoemission electron microscope with x-ray magnetic circular dichroism (XMCD-PEEM) for magnetic contrast to observe single magnet states of the ASI ensembles\citep{Aballe2015}.
An in-plane, bi-axial quadrupole magnet with two pairs of coils and a split 2D-yoke provides astroid clocking fields\citep{Foerster2016}.
The signal at the Fe \ce{L_3} edge is exploited for ferromagnetic XMCD contrast. 

The orientation of the ASI ensembles, applied magnetic fields, and XMCD contrast is carefully selected.
Samples are mounted with top and bottom ensemble edges parallel to the synchrotron light, with each nanomagnetic element oriented \SI{\pm45}{\degree} to the light.
This orientation guarantees balanced magnetic contrast for nanomagnets of both sublattices $L_a$ and $L_b$.
The in-plane field direction is given relative to the incoming x-ray illumination, with angle values increasing counter-clockwise. 
Consequently, the field directions and ensemble orientation align with the illustration in \cref{fig:astroid-clocking}, with an added light axis (providing magnetic contrast) parallel to the $h_x$-axis. 

The general experimental procedure is to initialize the ASI system, then apply clock protocols interspersed with magnetic imaging. 
We initialize the system by applying a strong, polarizing magnetic field (\SI{72}{\milli\tesla} along \SI{180}{\degree}), followed by two smaller fields, (\SI{18}{\milli\tesla} along \SI{0}{\degree} and \SI{3.5}{\milli\tesla} along \SI{180}{\degree}) to demagnetize the yoke. 
For the bipolar clocking, however, the initial field strength is \SI{82}{\milli\tesla}.
The difference in field strength is due to observed differences in the ensemble coercivity.
Successful initialization is confirmed by imaging a fully polarized ensemble (fully bright contrast (leftwards), as in snapshot $t=0$ of \cref{fig:all-AB-snapshots-1}) and the absence of remaining image translation in the PEEM (indicating a demagnetized yoke). 

After initialization, we perform steps of the clock protocols by alternating the application of clock pulses $A$, $B$, $a$ or $b$.
Each \textit{step} of a clock protocol comprises at least one \textit{clock pulse} (ramping the applied field to $\vec{H}_i$, holding the max field value, ramping down to zero applied field), and a magnetic contrast image acquisition. 
The value of $H$ that defines the $\vec{H}_i$ magnitudes is \SI{62}{\milli\tesla} for the unipolar clocking, and \SI{75}{\milli\tesla} for the bipolar clocking. 
After applying the first cycle of a clock protocol, before imaging, we shift the image, using the electron microscope optics, to re-center the ensemble, compensating for a small remanent magnetization in the yoke.
We carry out multiple cycles, each consisting of applying clock pulses and capturing an image, while maintaining the same image shift throughout.

In addition to the growth and reversal protocols, we conduct a control experiment by applying repeated clock pulses of $A$ and $B$ separately.

\subsection*{\label[methods]{sec:methods-flatspin}flatspin simulations}

Numerical simulations were done using flatspin, a large-scale ASI simulator \citep{Flatspin2022}.
flatspin approximates each nanomagnet as a point dipole with position $\vec{r}_i$ and orientation $\theta_i$.
Each dipole then has two possible magnetization directions along $\theta_i$, i.e., a binary macrospin $s_i \in \{-1, +1\}$.

Each spin $i$ is influenced by a total field
$\vec{h}_i = \vec{h}_\text{dip}^{(i)} + \vec{h}_\text{ext}^{(i)} + \vec{h}_\text{th}^{(i)},$ 
where $\vec{h}_\text{dip}^{(i)}$ is the total dipolar field from neighboring magnets, $\vec{h}_\text{ext}^{(i)}$ is a global or local external field, and $\vec{h}_\text{th}^{(i)}$ is a stochastic magnetic field representing thermal fluctuations in each magnetic element.
The total dipolar field is given by the magnetic dipole-dipole interaction,
\begin{equation}
\vec{h}_\text{dip}^{(i)} = \alpha \sum_{j \ne i}\frac{3\vec{r}_{ij}(\vec{m}_j \cdot \vec{r}_{ij})}{\lvert\vec{r}_{ij}\rvert^5} - \frac{\vec{m}_j}{\lvert\vec{r}_{ij}\rvert^3},
\end{equation}
where $\vec{r}_{ij}=\vec{r}_i-\vec{r}_j$ is the distance vector from spin $i$ to $j$, and $\alpha$ scales the dipolar coupling strength between spins.
The coupling strength $\alpha$ is given by $\alpha = \frac{\mu_0M}{4\pi a^3}$, where $a$ is the lattice spacing, $M$ is the net magnetic moment of a single magnet, and $\mu_0$ is the vacuum permeability.

Nanomagnet switching (magnetization reversal) occurs if the total field is directed against the current magnetization $\vec{m}_i$ and the magnitude of the field exceeds the coercive field $h_\text{c}$.
flatspin employs a generalized Stoner-Wohlfarth model, where $h_\text{c}$ depends on the angle of the total field $\vec{h}_i$ with respect to the magnet orientation.
Associated with each magnet is a switching astroid, which describes $h_\text{c}$ in terms of the parallel (easy axis) and perpendicular (hard axis) component of the total field, $\vec{h}_\parallel$ and $\vec{h}_\perp$.
The shape of the switching astroid is described by the equation
\begin{equation} \label{eq:esw}
    \left(\frac{h_{\parallel}}{b h_k}\right)^{2/\gamma} + 
    \left(\frac{h_{\perp}}{c h_k}\right)^{2/\beta} = 1,
\end{equation}
where $h_k$ denotes the coercive field along the hard axis.
The parameters $b$, $c$, $\beta$, and $\gamma$ adjust the shape of the astroid: $b$ and $c$ define the height and width, respectively, while $\beta$ and $\gamma$ adjust the curvature of the astroid at the easy and hard axis, respectively. 
Astroid parameters are typically tuned to obtain a shape that agrees with results from micromagnetic simulations.

Fabrication imperfections are modelled as variation in the coercive fields $h_{k}^{(i)}$, which are sampled from a normal distribution $\mathcal{N}(h_{k}, \sigma)$, where $\sigma = k_\text{disorder} \cdot h_k$ and $k_\text{disorder}$ is a user-defined parameter.

Dynamics are modeled using a deterministic single spin flip strategy.
At each simulation step, the total magnetic field $\vec{h}_i$ is calculated.
Next, we obtain a list of spins that \textit{may} flip, according to the switching astroid.
Finally, the spin which is \emph{furthest outside its switching astroid} is flipped.
The dipolar fields are recalculated after every spin flip, and the above process is repeated until there are no more flippable spins. 
This relaxation process is performed with constant external and thermal fields.

In this work, a global external field is used ($\vec{h}_\text{ext}^{(i)} = \vec{h}_\text{ext}$), and thermal fluctuations are assumed to be negligible ($\vec{h}_\text{th}^{(i)} = 0$).

The coupling strength $\alpha=0.0013$ was estimated to match the experimental results from the $50\times50$ fabricated pinwheel sample (see \cref{sec:methods-sample-fabrication}).
The value of $\alpha=0.0013$ is lower than predicted by theory ($\alpha \approx 0.0025$), which is likely due to demagnetizing oxidation of the permalloy.
A partially oxidized nanomagnet will have a reduced magnetic moment and a smaller effective size as the surface layer is no longer ferromagnetic.
The smaller $30\times30$ sample used in \cref{fig:aAbB-series} had a slightly larger magnet spacing and $\alpha=0.0012$ was used in this case.

For the simulation studies, a field strength $H = \SI{76.5}{\milli\tesla}$ and no disorder was used.
Simulations accompanying the experimental results used a slightly lower field strength of $H = \SI{75.8}{\milli\tesla}$ for \cref{fig:ab-series} and $H = \SI{75.9}{\milli\tesla}$ for \cref{fig:aAbB-series}.

Switching parameters were estimated from micromagnetic simulations of a \qtyproduct{220x80x25}{\nano\meter} stadium magnet using mumax\citep{Mumax2014}, namely $h_k=\SI{0.2}{\tesla}$, $b=0.38$, $c=1$, $\beta=1.3$, and $\gamma=3.6$
Other parameters include $k_\text{disorder}=4\%$ and a neighbor distance of $10$.

\backmatter

\setcounter{section}{0}
\setcounter{subsection}{0}
\setcounter{figure}{0}
\renewcommand*{\thesubsection}{S\arabic{subsection}}

\newpage
\section*{\label[supsection]{sec:supplementary}Supplementary information}

\subsection{\label[supsection]{sec:supplementary-neighbor-interactions}Neighborhood interactions}

\begin{figure*}
    \includegraphics[width=\textwidth]{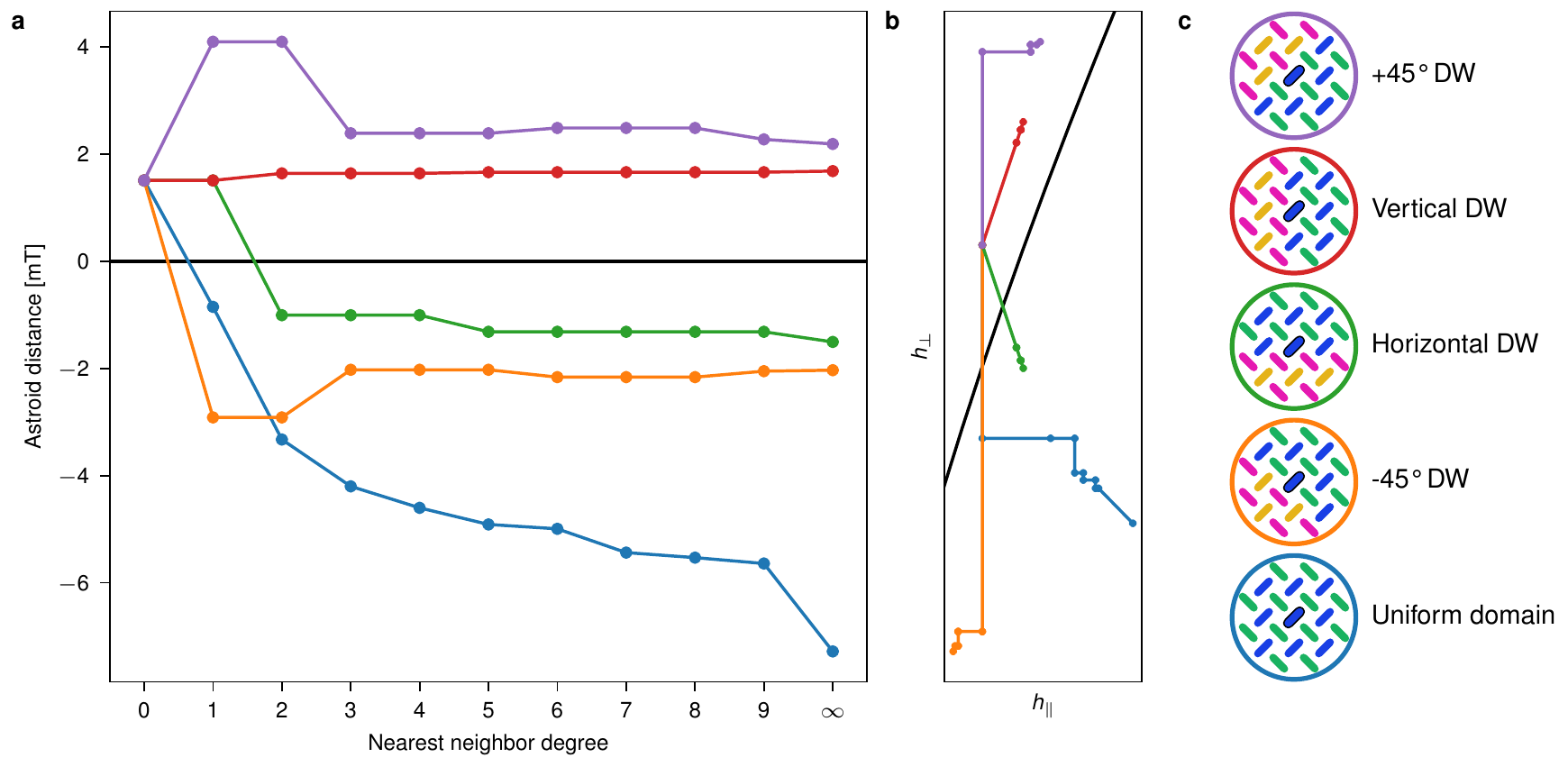}
    \caption{\label[supfigure]{fig:astroid-distance}%
    Minimum distance to the astroid edge as the neighborhood is increased, for the highlighted blue magnet in the center of each scenario in \textbf{c}.
    In all cases, the clock field $\vec{H}_A$ is applied.
    \textbf{a}, distance to the astroid for the highlighted magnet, as the neighborhood is increased when calculating the dipolar fields.
    \textbf{b}, shows a trace of the position within the astroid as the neighborhood is increased. Note that the scenarios all start at the same point (no neighbors), then diverge. 
    }
\end{figure*}

Here we analyse what type of neighbor interactions causes switching to occur selectively along the vertical and $+45\degree$ domain walls.
We consider five different prototype cases shown in  \cref{fig:astroid-distance}c: a uniform blue/green (leftwards) domain, and two domains separated by horizontal, vertical, and $\pm45\degree$ domain walls (DWs).
Within each prototype case, the subject of study is the highlighted blue magnet in the center.
The circled insets in the figure show only a limited neighborhood in the center of a larger $50\times50$ system which is initialized according to each prototype case.

\cref{fig:astroid-distance}a plots the distance to the astroid for the center magnet, as the number of neighbors are increased when calculating the dipolar fields.
In other words, we compute the total dipolar field from all magnets within a radius of the Nth nearest neighbor (NN).
After adding the total dipolar field to the external clock field $H_A$, the shortest distance to the astroid is calculated.
We define astroid distance as positive outside the astroid and negative inside.

Astroid distance is plotted for each of the five prototype cases in \cref{fig:astroid-distance}c.
With zero neighbors, and hence no dipolar fields, all five cases start at the same point outside the astroid.
As the first NNs are included, the cases split into four: the uniform domain and the $-45\degree$ domain wall enter the astroid.
In other words, the dipolar fields from the first NNs stabilize and prevent switching in these two cases.
Including also the second NNs causes the horizontal domain wall to enter the astroid. 
Horizontal domain walls are hence stabilized by 2nd NN interactions.
For the horizontal and $-45\degree$ domain walls, astroid distance does not change significantly as the neighborhood is increased further.
For the uniform domain, however, astroid distance increases further as the number of NNs are increased, with significant stabilizing interactions also beyond 9NNs.

Next, we consider the two cases where switching \emph{does} occur, namely the vertical and $+45\degree$ domain walls.
Somewhat curious, the astroid distance for the vertical domain wall appears to stay nearly constant across all NNs.
The $+45\degree$ domain walls travel further outside the astroid due to 1st NN interactions, then the 3rd NN interactions bring it closer to the astroid again, after which it remains at a near-constant distance.

\cref{fig:astroid-distance}b shows a trace of the location within the astroid as the NNs are increased.
For the vertical domain wall (red line), there is indeed movement due to dipolar interactions, but the movement is exclusively \emph{parallel} to the astroid edge.
Hence, the astroid distance in this case remains constant.
For the $+45\degree$ domain wall (purple line), the movement is purely in the perpendicular ($h_\perp$) direction for the 1st NN interactions, then purely parallel ($h_\parallel$) from the 3rd NN fields.

\begin{figure*}
    \includegraphics[width=\textwidth]{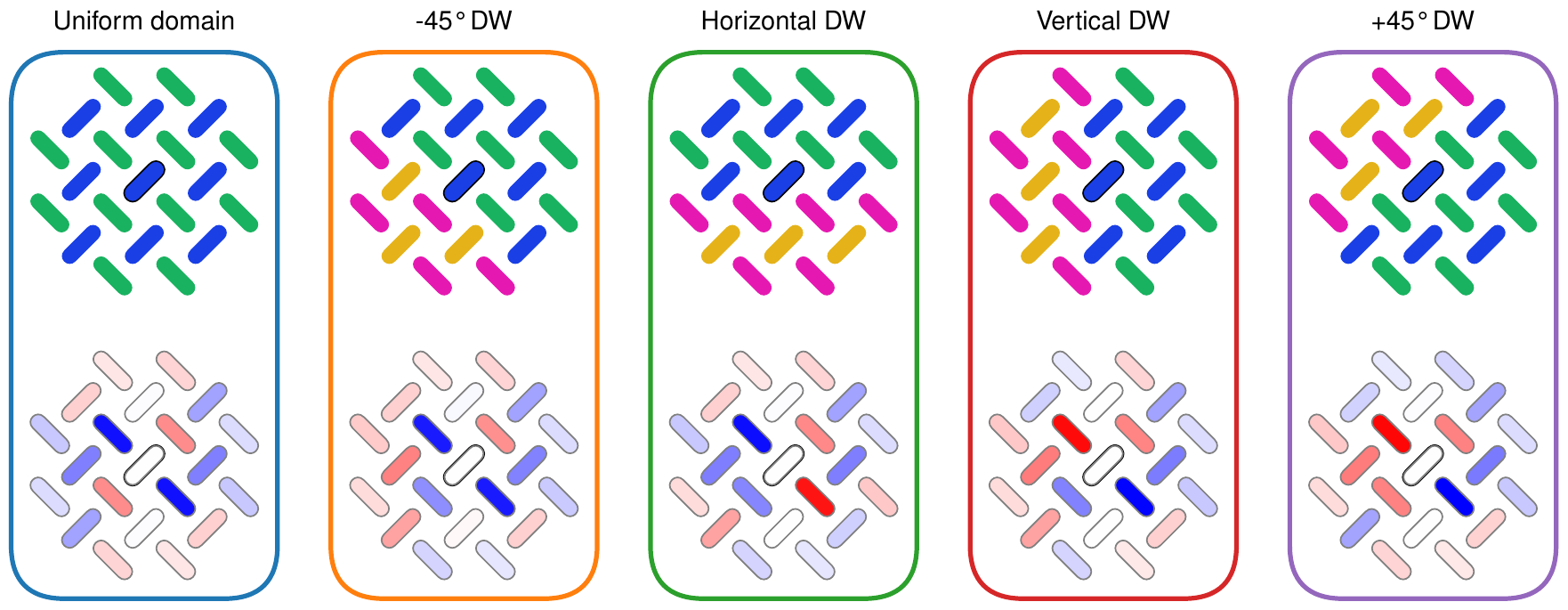}
    \caption{\label[supfigure]{fig:astroid-distance-neighbors}%
    Neighborhood influence with respect to center magnet. 
    Each scenario depicts the magnetization state (top) and the corresponding influence of the neighbors (bottom). 
    Stronger red signifies that the magnet is biasing the center magnet \textit{towards} switching, and stronger blue signifies that the magnet is biasing the center magnet \textit{away} from switching.
    }
\end{figure*}

An even more detailed picture is provided in \cref{fig:astroid-distance-neighbors}, where each neighbor magnet is colored according to the contribution of its dipolar field.
Specifically, a magnet is colored red (blue) if its dipolar field pushes the center magnet further out of (into) the astroid.
The shade of red (blue) represents how much the dipolar field contributes to promote (prevent) switching of the center magnet.
A magnet is colored white if its dipolar field has no contribution on the resulting astroid distance.

As can be seen in \cref{fig:astroid-distance-neighbors}, the neighborhood in the uniform domain is dominated by magnets that prevent switching (colored blue), with the highest contribution from the first NNs along the hard axis of the center magnet.
The same subset of the NNs are also the primary stabilizing force of the $-45\degree$ DW.
For the horizontal DW case, the dipolar fields from the first NNs cancel out, and it is the second NNs that prevent switching.

For the vertial DW, there is an apparent symmetry between neighbors that prevent and promote switching.
As a result the vertical DW is not stabilized and hence easily switched.
We saw earlier how this is because the dipolar fields are directed parallel to the astroid edge.
The $+45\degree$ DW is the least stable, where 3/4 of the first NNs promote switching (colored red).

\subsection{Growth and reversal in bipolar clocking\label[supsection]{sec:supplementary-bipolar-process}}
\begin{figure}[h]
    \includegraphics[width=\textwidth]{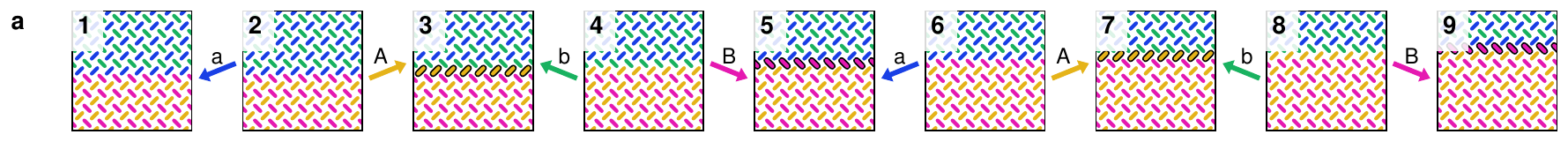}
    \includegraphics[width=\textwidth]{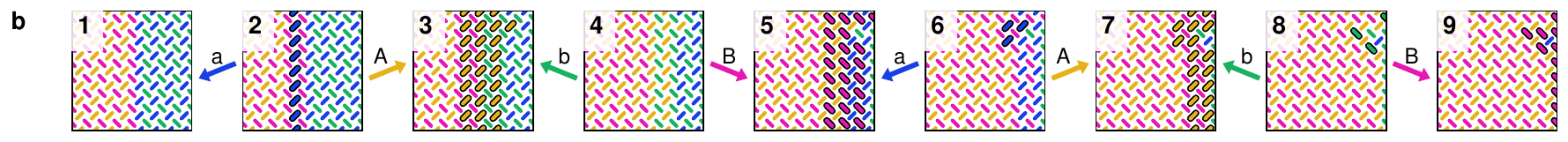}
    \includegraphics[width=\textwidth]{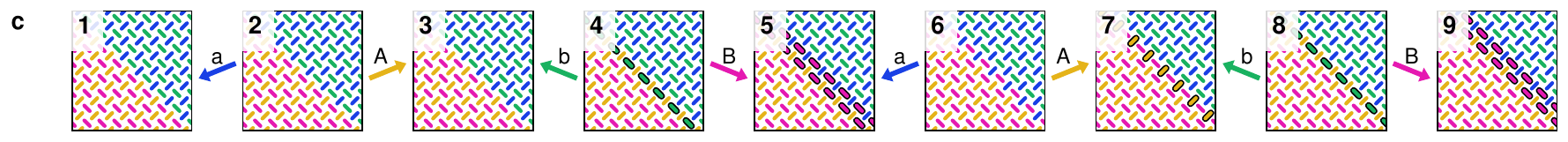}
    \includegraphics[width=\textwidth]{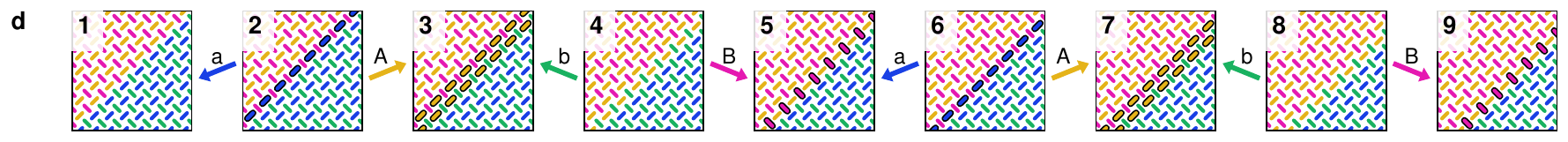}
    \caption{\label[supfigure]{fig:aAbB-clocking-dw}%
    Bipolar $aAbB$ clocking of four types of domain walls in pinwheel ASI: \textbf{a}, horizontal DW, \textbf{b}, vertical DW, \textbf{c}, $-45\degree$ DW and \textbf{d}, $+45\degree$ DW. 
    Each domain wall is initialized to fill the whole $50\times50$ system from edge to edge.
    Each snapshot shows a zoomed-in view of the system, at different points during a clock protocol.
    (1) shows the initial state.
    (2-12) show the state during $aAbB$ clocking.
    Magnets that change state between snapshots are highlighted by a solid black outline.
    }
\end{figure}

During bipolar clocking, domain growth and reversal in a single clock cycle can be observed for several domain wall configurations.
\cref{fig:aAbB-clocking-dw} shows the time evolution of different types of domain walls, subject to $aAbB$ clocking.
A straightforward example of simultaneous growth and reversal can be seen in \cref{fig:aAbB-clocking-dw}d, which shows a $+45\degree$ domain wall.
Notice that the first clock pulse $a$ moves the domain wall one step towards the left, and hence a reversal of the orange/pink domain.
However, the subsequent $A$ pulse immediately undoes this change \emph{and} moves the domain wall another step towards the right, advancing the domain wall a total of two layers of the sublattice $L_a$ (orange magnets).
Next, the $b$ pulse has no effect, since the pink magnets along the domain wall are stabilized by the dipolar fields from their neighbors.
Finally, the $B$ pulse moves the domain another step towards the right, flipping the next layer of magnets from sublattice $L_b$ (from green to pink).
As can be seen, the result is an apparent growth of the orange/pink domain by a single layer along the domain wall.
The other domain wall cases in \cref{fig:aAbB-clocking-dw} also show simultaneous growth and reversal, but are not discussed in further detail.

There is an apparent competition between growth and reversal.
For the $+45\degree$ domain wall discussed earlier, the competition seems to favor growth.
However, the situation strongly depends on the particular shape of the domain.
\cref{fig:aAbB-clocking} (main text) shows the time evolution of a hexagonal domain subject to $aAbB$ clocking.
As can be seen, the domain both grows horizontally and reverses vertically, and hence gradually changes shape over time.
Since vertical domain reversal depends on the curvature of the domain, the process will stop when the domain grows too wide.
The domain will continue to grow horizontally, as horizontal domain growth is not dependent on curvature.
As a result, domain growth seems to out-compete reversal in this case.
The end result is an apparent tendency towards horizontally elongated domains.

\begin{figure}[h]
    \centering
    \includegraphics[width=1.0\textwidth]{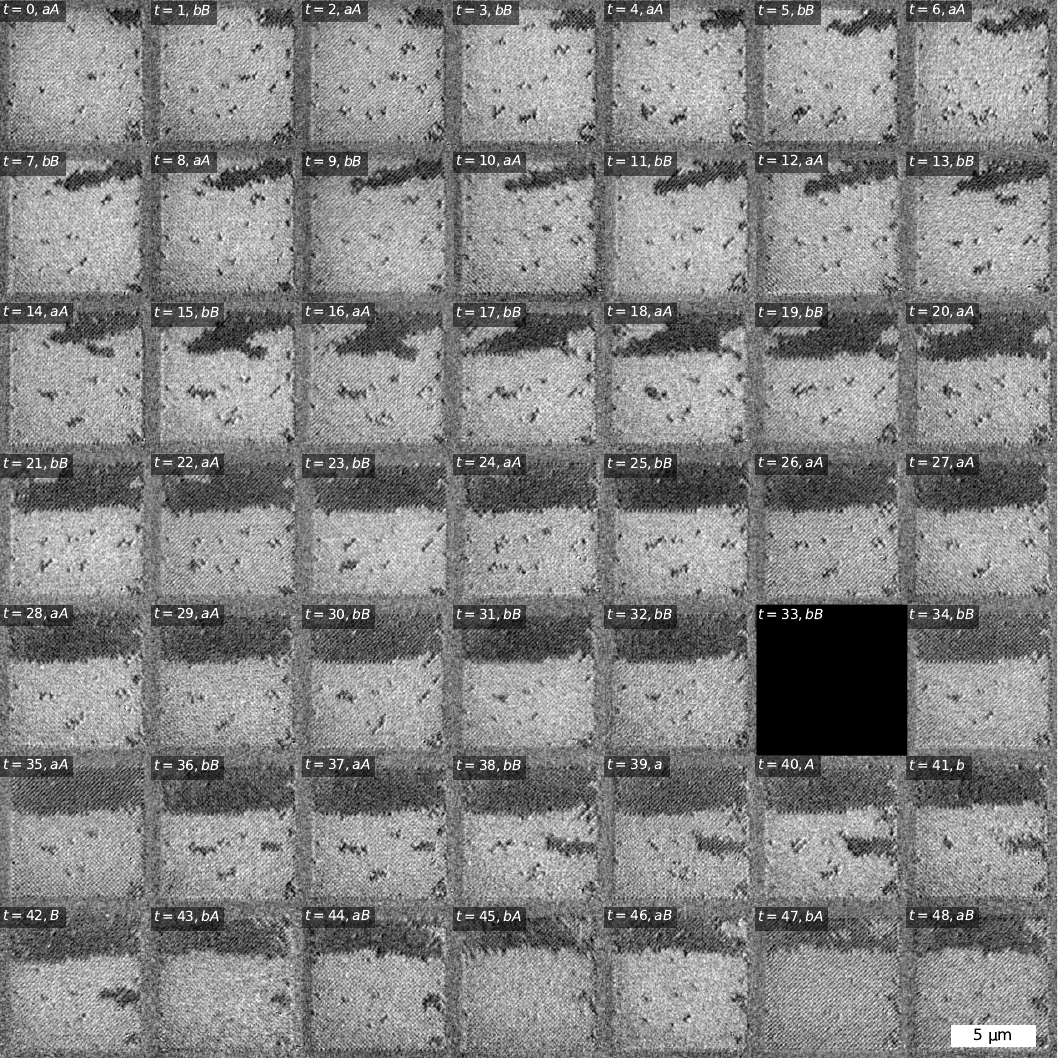}
    \caption{\label[supfigure]{fig:all-DACB-snapshots}%
    XMCD-PEEM images of all steps from the bipolar clock protocol series. 
    The time starts at $t=0$, and is incremented by 1 for each image, with clock pulses indicated by the labels. 
    The black (rightwards) domains grow and change shape as the $aAbB$ protocol is applied. 
    There are two control series where $aA$ and $bB$ are applied, where no change occurs. 
    Note that there is missing data for $t=33$, but the ensemble was still subjected to the clock pulses.
    At $t=39$ we image after each single clock pulse.
    From $t=42$ the reverse protocol $BbAa$ is applied, and the black (rightwards) domains shrink.
    Note that during the reversal protocol we still image after a final $A$ or $B$ pulse, in order to keep a constant image shift.
    }
\end{figure}

\clearpage
\subsection{Videos}

\renewcommand\figurename{Video}%
\setcounter{figure}{0}
    
\begin{figure}[h]
    \centering
    \includegraphics[width=.5\textwidth]{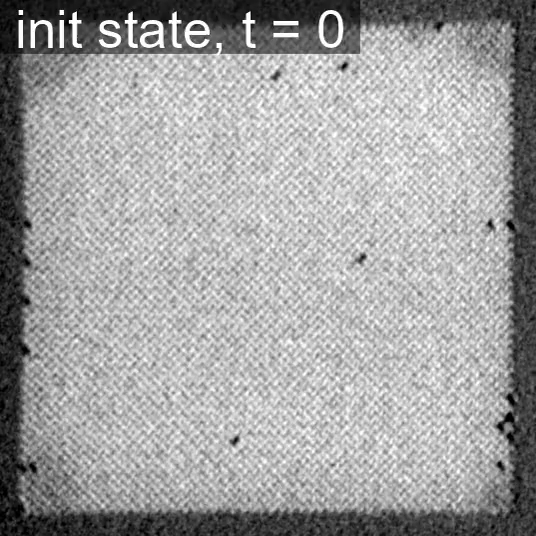}
    \caption{\label[supvideo]{vid:unipolar-peem}%
    Video showing all XMCD-PEEM images from the unipolar clock protocol series. 
    Each frame is labeled by the clock pulse(s) preceding it.
    The series consists of growth, reversal, re-initialization, control experiment and a second phase of growth.
    }
\end{figure}

\begin{figure}[h]
    \centering
    \includegraphics[width=.5\textwidth]{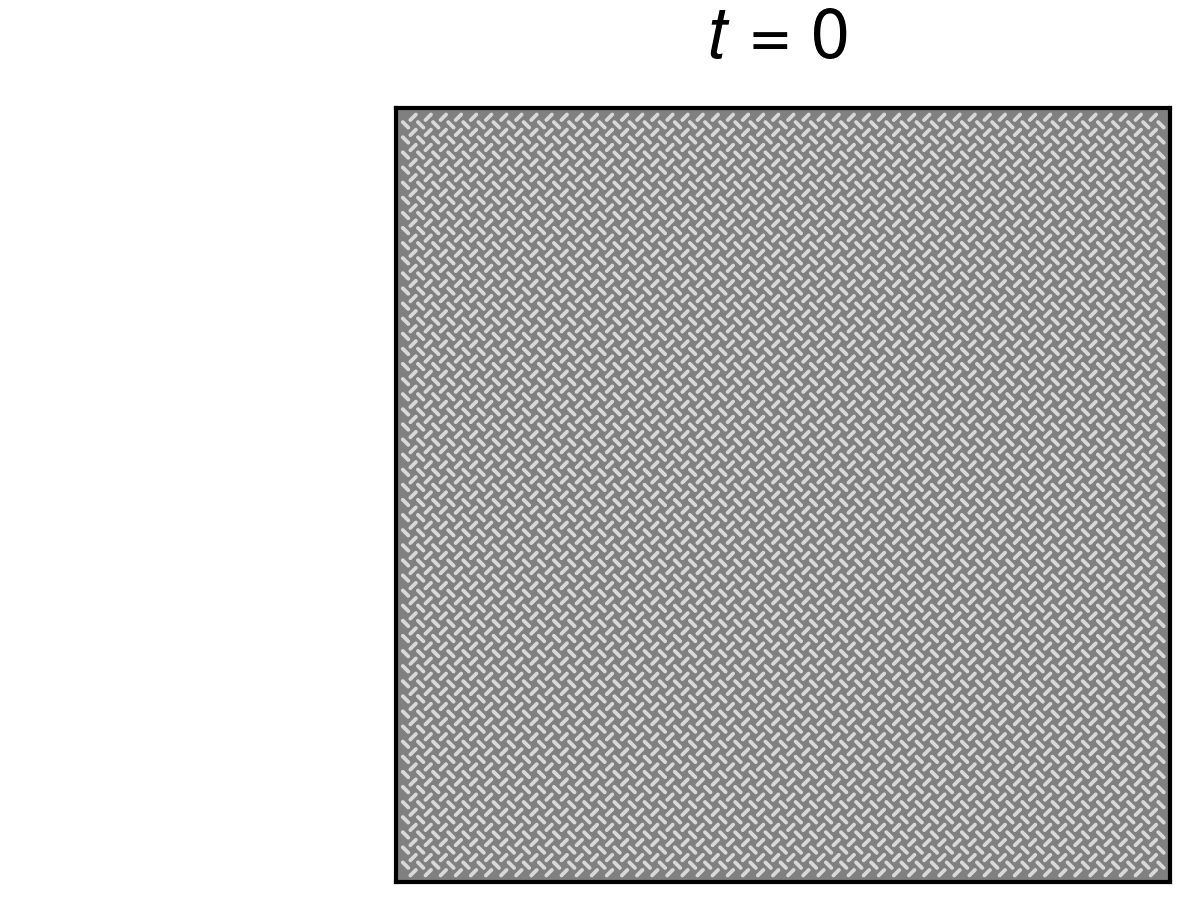}
    \caption{\label[supvideo]{vid:unipolar-sim}%
    Video showing flatspin simulation of the unipolar clock protocol series. 
    Each frame depicts the clock pulse preceding it.
    The series consists of growth, reversal, re-initialization and control experiment.
    }
\end{figure}

\begin{figure}[p]
    \centering
    \includegraphics[width=.5\textwidth]{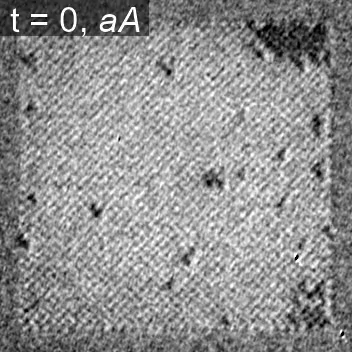}
    \caption{\label[supvideo]{vid:bipolar-peem}%
    Video showing all XMCD-PEEM images from the bipolar clock protocol series. 
    Each frame is labeled by the clock pulse(s) preceding it.
    The series consists of growth, control experiment and reversal.
    }
\end{figure}

\begin{figure}[p]
    \centering
    \includegraphics[width=.5\textwidth]{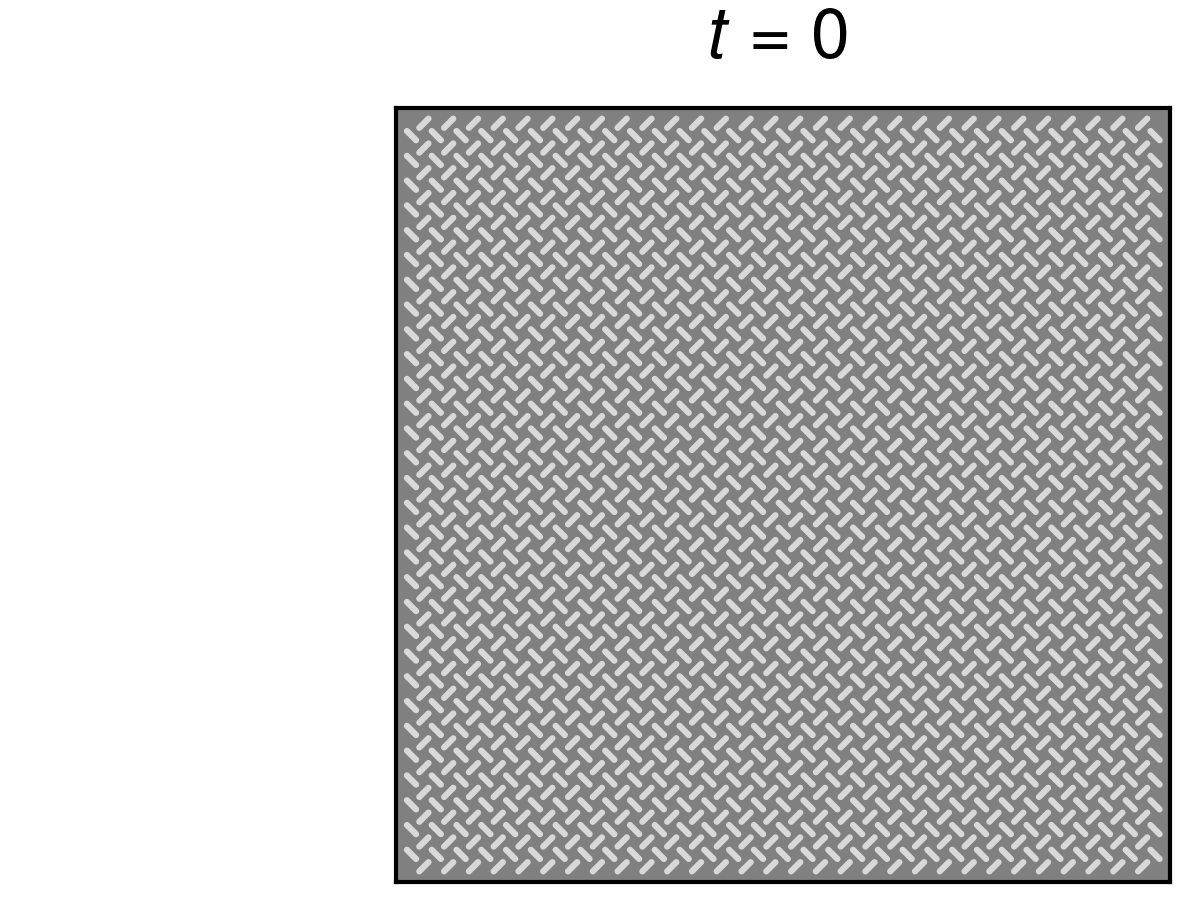}
    \caption{\label[supvideo]{vid:bipolar-sim}%
    Video showing flatspin simulation of the bipolar clock protocol series. 
    Each frame depicts the clock pulse preceding it.
    The series consists of growth, control experiment and reversal.
    }
\end{figure}

\clearpage
\section*{Acknowledgments}
These experiments were performed at the CIRCE beamline at ALBA Synchrotron with the collaboration of ALBA staff. 

This work was funded in part by the Norwegian Research Council through the IKTPLUSS project SOCRATES (Grant no. 270961) and the TEKNOKONVERGENS project SPrINTER (Grant No. 331821), and in part by the EU FET-Open RIA project SpinENGINE (Grant No. 861618).

The Research Council of Norway is acknowledged for the support to the Norwegian Micro- and Nano-Fabrication Facility, NorFab, project number 295864.
Simulations were executed on the NTNU EPIC compute cluster\cite{Epic2019}.

MAN, MF and MWK acknowledge funding from MCIN through grant number PID2021-122980OB-C54 and MWK also acknowledges support through Marie Sklodowska-Curie grant agreement No. 754397 (DOC-FAM) from EU Horizon 2020.

\bibliography{bibliography}%

\end{document}